\documentclass[aps,pre,twocolumn,superscriptaddress]{revtex4-2}
\usepackage{amssymb}
\usepackage{amsmath,bm}
\usepackage{graphicx,color}
\usepackage{bbm}
\usepackage[bottom]{footmisc}
\usepackage{multirow}
\usepackage{xcolor}
\usepackage{xpatch} % <<<<<<<<<<<<<
\usepackage[normalem]{ulem} %for strikethrough using \sout
\renewcommand{\eqref}[1]{Eq.(\ref{#1})}
\newcommand{\secref}[1]{Sec.(\ref{#1})}
\newcommand{\figref}[1]{Fig.~\ref{#1}}
\newcommand{\appref}[1]{App.~\ref{#1}}

\newcommand{\rin}{r_\mathrm{in}}
\newcommand{\rout}{r_\mathrm{out}}

\newcommand{\dx}{\dif x}
\newcommand{\gbar}{\bar{g}}
\newcommand{\nablabar}{\bar{\nabla}}

\newcommand{\Kbar}{\bar{K}}

\newcommand{\brk}[1]{\left(#1\right)}

\newcommand{\A}{\mathcal{A}}
\newcommand{\dif}{\mathrm{d}}

\newcommand{\M}{\mathcal{M}}
\newcommand{\N}{\mathcal{N}}

\newcommand{\ep}{\varepsilon_0}
\newcommand{\E}{\mathbf{E}}
\newcommand{\Pol}{\mathbf{P}}
\newcommand{\U}{\mathcal{U}}
\newcommand{\W}{\mathcal{W}}

\newcommand{\deltan}{\delta^{(n)}}
\newcommand{\xvec}{\mathbf{x}}
\newcommand{\uel}{u^\mathrm{el}}
\newcommand{\Deltabar}{\bar{\Delta}}
\newcommand{\sigmael}{\sigma_\mathrm{el}}
\newcommand{\qvec}{\mathbf{q}}

\newcommand{\pvec}{\mathbf{p}}

\newcommand{\ellp}{\ell_{P}}

\newcommand{\Noemie}[1]{\noindent \color{cyan}  #1\normalcolor}

\begin{document}
\title{Geometric Theory of Mechanical Screening in two-dimensional solids}
\author{Noemie Livne}

\author{Amit Schiller}
% \affiliation{Racah Institute of Physics, The Hebrew University of Jerusalem, Jerusalem, Israel 9190}
%\author{Oran Szachter}
% \affiliation{Racah Institute of Physics, The Hebrew University of Jerusalem, Jerusalem, Israel 9190}
\author{Michael Moshe}
\email{michael.moshe@mail.huji.ac.il}
\affiliation{Racah Institute of Physics, The Hebrew University of Jerusalem, Jerusalem, Israel 9190}

\begin{abstract}
Holes in mechanical metamaterials, quasi-localized plastic events in amorphous solids, and {bound} dislocations in a hexatic matter are different mechanisms of generic stress relaxation in solids. 
Regardless of the specific mechanism, these and other local stress relaxation modes are quadrupolar in nature, forming the foundation for stress screening in solids, similar to polarization fields in electrostatic media. 
We propose a geometric theory for stress screening in generalized solids based on this observation. 
The theory includes a hierarchy of screening modes, each characterized by internal length scales, and is partially analogous to theories of electrostatic screening such as dielectrics and Debye-Hückel theory. 
Additionally, our formalism suggests that the hexatic phase, traditionally defined by structural properties, can also be defined by mechanical properties and may exist in amorphous materials.
\end{abstract}
\maketitle
%%%%%%%%%%%%%%%%%%%%%%%%%%%%%%%%%%
\section{Introduction}
\label{intro}
The concept of screening, which refers to the reduction of energy density through a material's local responses, is central to many physical systems.  
Examples include dielectrics and ionic liquids, in which induced dipolar or monopolar charge densities respond to the background electric field.
As a result, the effective electric field is modified either quantitatively or qualitatively \cite{LL-EDCM}.
Previous research has successfully applied the concept of screening to mechanical systems. For instance, the onset of buckling in 2D defective membranes has been interpreted as the screening of structural defects by curvature \cite{seung1988defects}. Additionally, studies have shown that mechanical stresses in curved self-assembled crystals can be screened through the nucleation of structural defects \cite{nelson2002defects, bausch2003grain, irvine2010pleats, irvine2012geometric}. 

The duality between curvature and defects as entities that screen and are screened is reflected in the first Föppl–von Kármán equation for the stress potential $\chi$
\begin{eqnarray}
	\frac{1}{Y} \Delta\Delta\chi = K_{D} - K_G \;.
	\label{eq:Airy}
\end{eqnarray}
In this equation, the Gaussian curvature of the actual deformed configuration is represented by $K_G$, and singular or distributed defects are represented by $K_D$ \cite{seung1988defects}. This equation demonstrates that when the curvature $K_G$ is fixed, stresses can be reduced by distributing defects through $K_D$, and vice versa.

\begin{figure*}
    \centering
    \includegraphics[width = \linewidth]{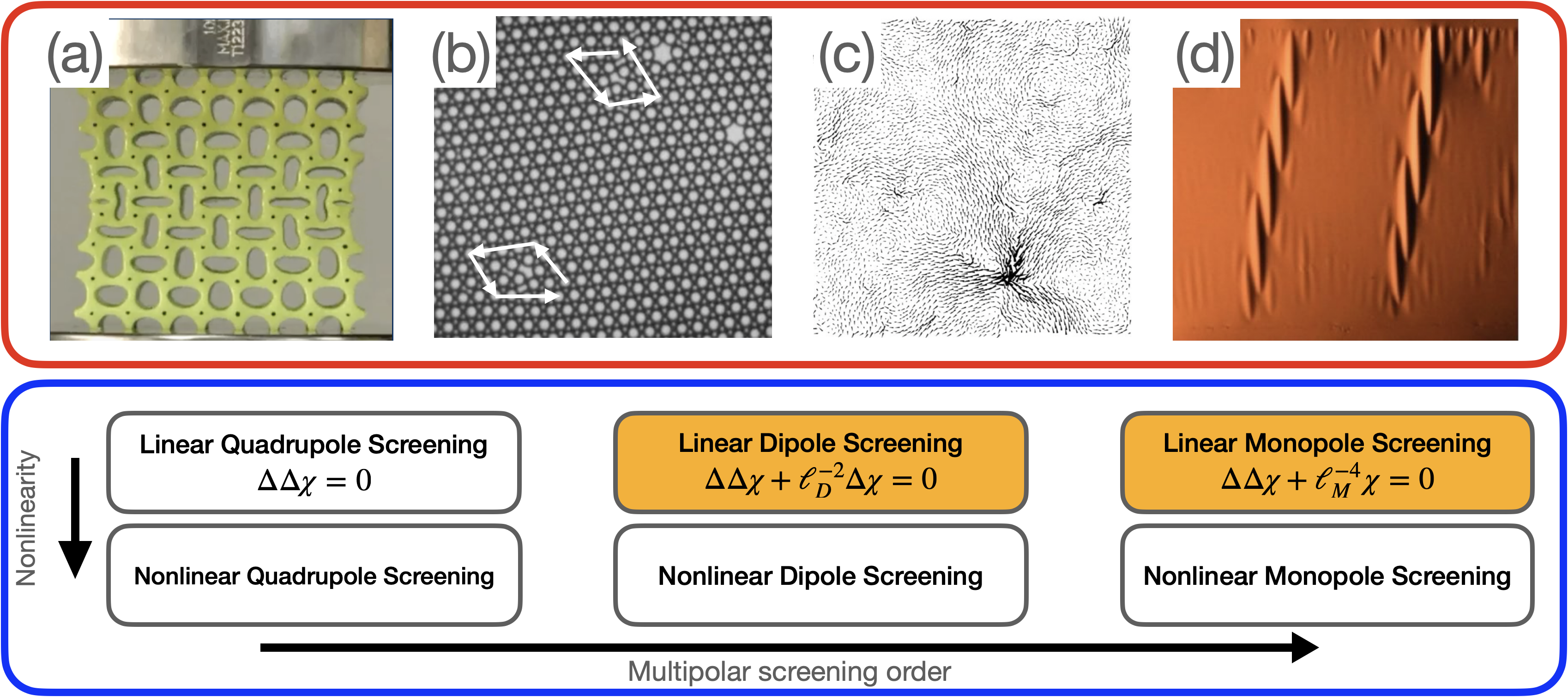}
    \caption{Mechanical Screening. Top panel -Stress relaxation mechanisms: (a) Nonlinear quadrupole screening in holey metamaterials, established in \cite{bar2020geometric}, (b) Screening by dipoles via dislocation unbinding in a 2d crystal bubble-raft model \cite{bragg1947dynamical,bowick2008bubble}, (c) Quadrupolar Eshelby plastic event in a model of amorphous solid, adapted with permission from \cite{MaloneyLemaitre06}, (d) Screening by local quadrupolar wrinkles, adapted with permission from \cite{timounay2020crumples}. 
    Bottom panel - diagrammatic description of the different screening modes. The linear and nonlinear quadrupole screening theory was established in \cite{bar2020geometric}. Here we focus on linear dipole and monopole screening theories, extending linear quadrupole screening in analogy with the extension of dielectrics to Debye-Hückel screening.}
    \label{fig:mechScreeningHierarchy}
\end{figure*}

Physical phenomena that can be explained by geometric screening include the shape of virus capsids \cite{lidmar2003virus} and defect patterns on curved colloidal crystals \cite{irvine2010pleats, bausch2003grain}. Another example is the theory of linear and nonlinear screening by imaginary quadrupoles, which was systematically derived to describe the emergent mechanics in Kirigami \cite{moshe2019nonlinear} and planar elastic meta-materials containing arrays of holes \cite{bar2020geometric}. In \figref{fig:mechScreeningHierarchy}(a) we demonstrate a state in which imaginary quadrupoles interact nonlinearly, leading to a spontaneous breaking of symmetry with an alternate pattern
\cite{bar2020geometric}.    

Previous works on mechanical screening have been largely influenced by an early discovery of mechanical screening within the statistical theory of 2D crystalline matter, which led to the concept of two-step melting of a solid through an intermediate hexatic phase to a liquid state
\cite{halperin1978theory,nelson1979dislocation}.
In this theory, the three phases are distinguished by their structural properties, and the transitions from solid to hexatic and hexatic to liquid correspond to a sequential destruction of translational and rotational quasi-long-range order.

From a mechanical perspective, the low, intermediate, and high temperature phases form elastic solids supplemented by thermally induced tightly-bounded dislocation-pairs, tightly bounded disclination-pairs (dislocations), and free disclinations, respectively. The free element in each phase forms a potential screening mechanism. In the intermediate hexatic phase, for example, dislocations can form in pairs and unbind to screen out external loads, and are the key mechanism behind its vanishing shear modulus and the screened interactions between disclinations \cite{halperin1978theory, nelson1979dislocation, zhai2019two,pretko2019crystal}. This is illustrated in \figref{fig:mechScreeningHierarchy}(b) where a bubble-raft model of a 2d crystalline matter shows the unbinding of dislocations due to external shear.

The ever-growing list of systems that contain screening mechanisms is not limited to ordered systems.
Examples include granular amorphous solids, where local quadrupolar particle rearrangements are induced in response to external loads \cite{MaloneyLemaitre06} (shown in  \figref{fig:mechScreeningHierarchy}(c)), epithelial tissue \cite{park2015unjamming,farhadifar2007influence}, and wrinkles and crumples in strongly confined thin sheets, where local out-of-plane deformations are also of quadrupolar nature \cite{timounay2020crumples, aharoni2017smectic} (shown in \figref{fig:mechScreeningHierarchy}(d)).  

Motivated by the wide range of screening mechanisms found in solids, a linear continuum theory was developed to describe various modes of screening in elastic materials \cite{SchreiberKeren2021}. 
Specifically, two distinct screening regimes were predicted: a quasi-elastic regime and an anomalous one. 
It was suggested  that a transition between these different screening modes can be achieved, 
for example, in granular solid by decreasing the confining pressure.

Indeed, the theory's predictions, including the emergence of anomalous mechanics, have been validated through a series of numerical and experimental studies on the mechanics of granular and glassy materials in both two and three dimensions \cite{lemaitre2021anomalous,mondal2022experimental,kumar2022anomalous, bhowmik2022direct,charan2022Anomalous}. 
Despite its success in predicting the mechanics of granular and glassy materials, the theory presented in \cite{SchreiberKeren2021, lemaitre2021anomalous,mondal2022experimental,kumar2022anomalous, bhowmik2022direct,charan2022Anomalous} is derived based on ad hoc assumptions on the general nature of screening.
In addition, we identify three main drawbacks of the theory: (i) It is written in a specific coordinate system. (ii) It assumes a geometrically linearized strain measure. (iii) The analytic methods available within the current displacement-formulation are limited.

In this paper, we derive a hierarchy of screening theories from (geometric) first principles.
We address the limitations of previous theories by developing a covariant geometric formulation of screened elasticity. 
Our theory reveals three distinct screening regimes, controlled by quadrupole, dipole, and monopole screening mechanisms. 
Additionally, we develop a generalized Airy potential theory, in which the governing equations take different forms in each of the regimes
\begin{equation}
    \begin{aligned}
        &\frac{1}{\tilde Y}\Delta \Delta \chi  = \bar{K}^0  \quad &&\text{Quadrupole} \\
        &\frac{1}{\tilde Y}\Delta \Delta \chi + \frac{1}{ \tilde Y} \ellp^{-2} \Delta \chi  = \bar{K}^0  \quad &&\text{Dipole} \\
                &\frac{1}{ \tilde Y}\Delta \Delta \chi + \frac{1}{\tilde Y} \ell_M^{-4} \chi  = \bar{K}^0  \quad &&\text{Monopole}
    \end{aligned}
    \notag
\end{equation}
Our study demonstrates that the different  screening regimes are characterized by different length scales, $\ellp$ and $\ell_M$, which act as new moduli that extend classical elasticity. The theories of Dipole and Monopole screening predict non-affine deformations in response to uniform external loads and are expected to be relevant to any solid whose mechanics is controlled by local relaxation mechanisms, such as local rearrangements in amorphous solids, wrinkles in confined thin sheets, and T1 transitions in living cellular tissue.

The possible extensions of continuum mechanical screening are summarized in bottom panel of \figref{fig:mechScreeningHierarchy}. In this work we focus on the yellow-colored boxes representing linear dipole and monopole screenings, in which an unusual or anomalous mechanical behavior is predicted.

Our theory allows studying new problems that the non-geometric formulation in \cite{SchreiberKeren2021, lemaitre2021anomalous} could not address.
For example, we show that a monopole elastic charge screened by dipoles is mechanically equivalent to a disclination screened by dislocaitons in the Hexatic phase. Furthermore, we study how screened defects interact via the screening field. These and other predictions are proposed as test measurements for identifying mechanical screening. 
Surprisingly, the geometric approach to mechanical screening uncovered an explicit link between the mechanics of the hexatic phase within the theory of melting, and the mechanics of screened solids, even in the absence of underlying order.

The structure of this paper is as follows: 
We start with introducing an electrostatic analog in \secref{sec:ElAn},  where we derive electrostatic screening theories from energy functional minimization, an approach that is more natural when athermal mechanical systems are considered. In \secref{sec:GeometricScreening} we develop the general framework of geometric screening in elastic-like solids. In \secref{sec:Equilibrium} we derive equilibrium equations for the different screening modes, followed by the development of generalized screened Airy stress function approach in \secref{sec:Potential}.
In \secref{sec:Applications} we study the implications of mechanical screening on basic physical properties such as the Green's function associated with each screening mode, and the interactions between sources of stresses in the presence of screening.  In \secref{sec:Sum} we conclude by discussing the future road map towards a general theory of screening in solids.

\section{The electrostatic Analog}
\label{sec:ElAn}
\begin{figure}
    \centering
    \includegraphics[width = \columnwidth]{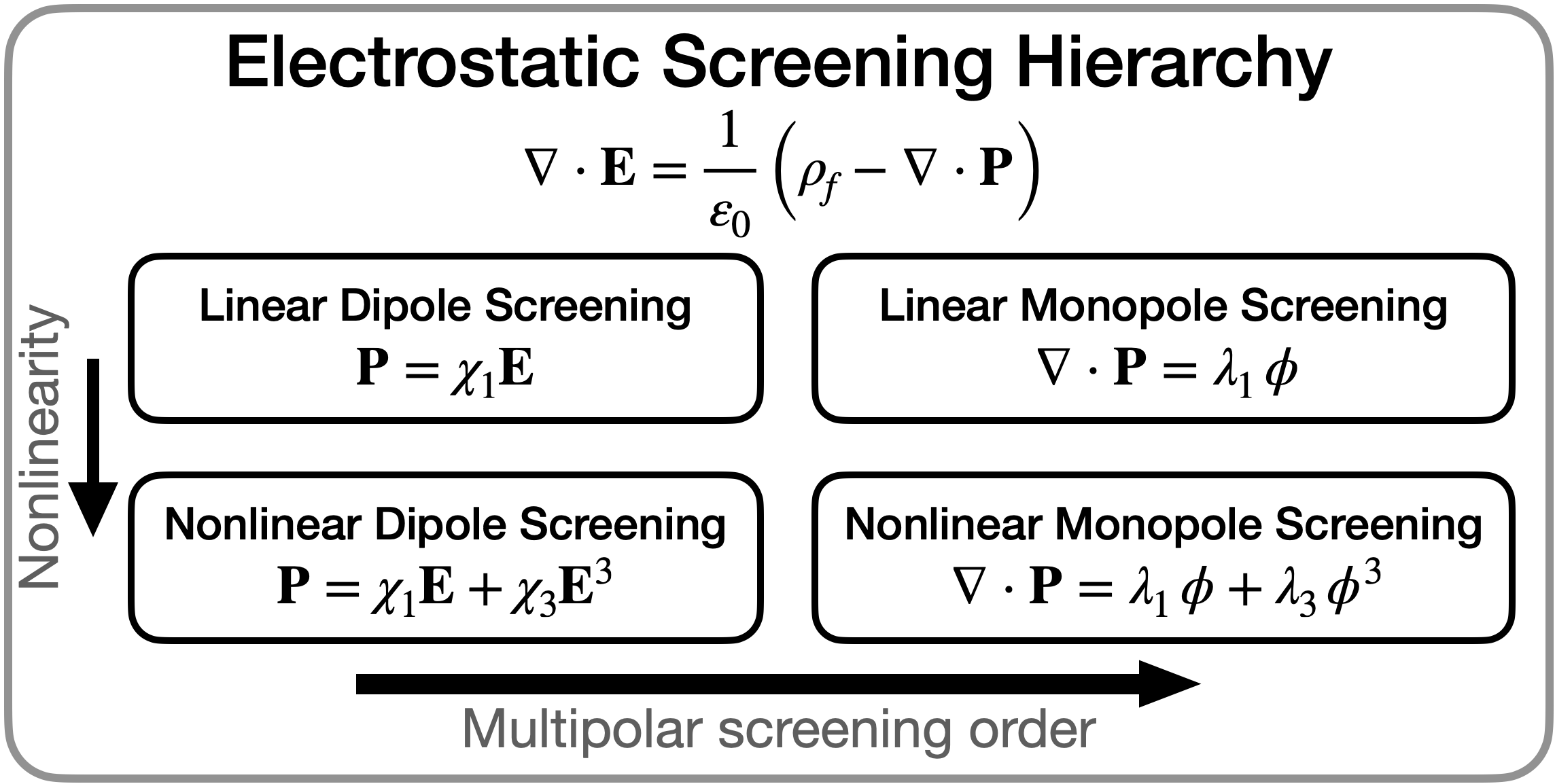}
    \caption{Diagrammatic representation of screening hierarchy in electrostatic media. The equation for the electric field depends on the induced polarization $P$ which depends on the electric field via a constitutive relation, illustrated here for each screening regime.}
    \label{fig:EMscreeningHierarchy}
\end{figure}

A familiar implementation of screening theory is within electrostatics of continuous media. {As such, we} find it instructive to start with the electrostatic analog and later implement the same ideas, with the necessary adjustments, to elastic solids. The main idea behind the analogy is the hierarchical structure of linear and nonlinear electrostatic screening as summarized in \figref{fig:EMscreeningHierarchy}.

The potential energy density stored in the electric field is $ \mathcal{U} = \tfrac{1}{2}\ep \E^2 $, and the work done on the system by assembling a charge density $\rho_f$ is $\mathcal{W} = \rho_f \phi$. The mechanical free energy in a domain $\M$ is therefore
\begin{equation}
	F =  \int_{\M} \left(\mathcal{U} - \mathcal{W}\right) \dif S = \int_{\M} \left( \frac{1}{2}\ep \E^2 - \rho_f \phi \right) \dif S  
	\label{eq:FreeEnergy}
\end{equation}
with $\E = -\nabla \phi$ the electric field derived from a potential, and $\ep$ the vacuum permittivity. {If the domain} $\M$ {is filled} with matter, atoms and molecules {may polarize} in response to electric field, {creating} electric dipoles that modify the electric field. 
At the continuum level the dipoles are described by the polarization density $\Pol$ \cite{LL-EDCM}. The self interaction energy of a dipole, or the work required for its nucleation, is material dependent and reflects the microscopic origin of the charge separation within the atom or the molecule. To account for this effect we note that the energetic cost is quadratic in the polarization, and that dipoles interact with each other via the total electric field, so
\begin{eqnarray}
		\U &=& \frac{1}{2}\ep \E^2 + \E \cdot \Pol \notag\\
		\W &=&  \frac{1}{2 \ep \chi_e} \Pol^2   + \rho_f \phi \;.
		\label{eq:dielectric}
\end{eqnarray}
Here $\chi_e$ is the electric susceptibility, and as before, $\U$ quantifies the energy stored in the electric field and $\W$ the work done on the system by assembling the monopole and dipole densities {$\rho_f$} and $ \Pol$.
The equilibrium equations are then
\begin{eqnarray}
	\Pol &=& \ep \chi_e \E \notag \\
	\nabla \cdot \E  &=& \frac{1}{\ep} \left( \rho_f - \nabla \cdot \Pol \right)
\end{eqnarray}
Upon substituting the first relation in the second we get
\begin{eqnarray}
	\nabla \cdot \E  &=& \frac{1}{\ep(1+\chi_e)}\rho_f  = \frac{1}{\varepsilon}\rho_f 
\end{eqnarray}

Thus, we see that the permittivity constant is renormalized by the induced dipoles. 
These equations are the basis for linear dielectrics. 

An important observation is that the form of $\W$ in \eqref{eq:dielectric} is not the most general one. 
Upon assuming that $\W$ is an analytic function of $\Pol$ and its derivative, the most general form that preserves the symmetries to rotations and translations is 
\begin{eqnarray}
	\W &=& \rho_f \phi + \frac{1}{2}\alpha_2 \Pol^2 + \frac{1}{24} \alpha_4 \Pol^4 + \dots \notag \\
	&+ & \frac{1}{2}\beta_2 (\nabla \cdot \Pol)^2 + \frac{1}{24}\beta_4 (\nabla \cdot \Pol)^4 + \dots \notag \\ 
	&+ & \frac{1}{2}\gamma_2 (\nabla \times \Pol)^2 + \frac{1}{24}\gamma_4 (\nabla \times \Pol)^4 + \dots \notag 
	\label{eq:Landau}
\end{eqnarray}
Within a linear theory, only three terms contribute, with nonzero $\alpha_2,\beta_2,\gamma_2 $, and perhaps additional terms in quadratic higher order derivatives. However, from a physical perspective, the interpretation of $\Pol$ as a polarization field\Noemie{,} together with the multipole expansion
\begin{eqnarray}
	\rho = \rho_f + \nabla \cdot \Pol + \nabla \nabla Q + \dots \Noemie{,}
\end{eqnarray}
imply that $\nabla\times\Pol$ does not contribute to the charge distribution. Hence, in electrostatic systems we expect $\gamma_2 = 0$.  For the same reason, higher order derivatives of $\Pol$ are irrelevant, leaving the general form
\begin{eqnarray}
	\W &=& \rho_f \phi + \frac{1}{2}\alpha_2 \Pol^2 +  \frac{1}{2}\beta_2 (\nabla \cdot \Pol)^2.\       
	\label{eq:Landau2}
\end{eqnarray}
The term proportional to $(\nabla \cdot \Pol)^2$ represents the nucleation cost associated with effective monopoles, created by non-uniformly distributed dipoles.
The two coefficients correspond to an inherent length scale $\ell \equiv \sqrt{\frac{\beta_2}{\alpha_2}}$. When compared with system size, 
the dielectric state corresponds to $\ell \ll L$. 
In the other limit, $L \ll \ell $, the term $\Pol^2$ is negligible, and
\eqref{eq:dielectric} takes the form
\begin{eqnarray}
	\U &=& \frac{1}{2}\ep \E^2 + \E \cdot \Pol \notag\\
	\W &=&  \frac{1}{2 \ell_0^2} (\nabla \cdot  \Pol)^2   + \rho_f \phi \;.
	\label{eq:debye}
\end{eqnarray}
Upon minimizing  $F =  \int_{\M} \left(\mathcal{U} - \mathcal{W}\right) \dif S $
%\eqref{eq:FreeEnergy} with those $\U$ and $\W$ 
the equilibrium equations are 
\begin{eqnarray}
	\nabla (\nabla \cdot \Pol ) &=& -\ep \ell_0^{-2} \E \notag \\
	\Delta \phi &=& -\frac{1}{\ep}\left(\rho_f - \nabla \cdot \Pol \right)\;.
	\label{eq:Debye}
\end{eqnarray}
The first equation can be written as 
\begin{eqnarray}
	\nabla (\nabla \cdot \Pol  - \ep \ell_0^{-2}  \phi) = 0
\end{eqnarray}
implying that the expression in brackets is constant which can be set to zero using the potential gauge freedom
\begin{eqnarray}
	\nabla \cdot \Pol  - \ep \ell_0^{-2}  \phi = 0 \;.
	\label{eq:LinearDebye}
\end{eqnarray}
Since the gauge is fixed, from this point onward we should no longer expect the equations to be invariant under gauge transformations.
Upon substituting \eqref{eq:LinearDebye} in \eqref{eq:Debye} we find
\begin{eqnarray}
	\Delta \phi  -   \ell_0^{-2}  \phi= -\frac{1}{\ep}\rho_f
	\label{eq:Helmholtz}
\end{eqnarray}
This is {the} Helmholtz equation from {the} Debye-Hückel theory, describing screening by mobile monopole charges in an ionic liquid.
We emphasize that in both dipole and monopole screenings, the fundamental fields with respect to which the energy is minimized are the electric potential and the polarization field. In the monopole screening case, the variation with respect to the polarization enforces the conservation of total charge.
\eqref{eq:Helmholtz} is traditionally derived from the Poisson-Boltzman equation using a detailed microscopic theory, which gives an explicit expression for the Debye-screening length $\ell_0 $  in terms of temperature, ionic strength etc.

Our minimization approach avoids the microscopic statistical picture, and thus provide no details on the parameter $\ell_0$. 
Despite this weakness, such an approach is advantageous in this work, since the systems we are interested in are mostly athermal and disordered.

\section{Pure and Screened Elasticity}
\label{sec:GeometricScreening}
One challenge in writing a screening theory for solids is the identification of the basic screening element, which arises naturally from a geometric approach to elasticity \cite{KMS2015}. In this formulation the reference state of a solid $\M$ is defined by the rest distances between material elements, and quantified by the reference metric $\gbar^0  $ via $\dif l_0^2 = \gbar^0_{\mu\nu} \dx^\mu \dx^\nu$. A configuration is described by the metric $g$, quantifying the actual (potentially deformed) distances between material elements given by $\dif l^2 = g_{\mu\nu} \dx^\mu \dx^\nu$. Contrary to the reference metric, the actual one is induced from an embedding $\phi:\mathcal{M}\to\mathbb{R}^2$ describing the material configuration with $g = \nabla \phi ^T \nabla \phi $. The strain is defined as the deviation of $g$ from its rest state $u = \tfrac12(g - \gbar^0)$.
A key property in this formulation is the curvature associated with the reference metric. A stress free configuration is available if the reference Gaussian curvatures $\Kbar^0  $ associated with $\gbar^0  $, vanishes. Therefore $\Kbar^0  $ is a measure of geometric incompatibility, and consequently for sources of residual stresses. 
Singular sources of stresses are described by singular $\Kbar^0$, exhibiting a natural multipolar hierarchy, as shown in Table \ref{tab:table1}.
\begin{table}[h]
	\begin{ruledtabular}
		\begin{tabular}{lcc}
			Type   &  $\bar{K}$ & Realization  \\ \hline
			Monopole & 
			$ m\, \delta(\xvec)$
			&
			Disclination
			\\ 
			Dipole & 
			$ \pvec\cdot\nabla\delta(\xvec)$ 
			&
			Dislocation
			\\ 
			Quadrupole & 
			$ (\nabla^T\cdot \qvec\cdot \nabla) \delta(\xvec)$ 
			&
			Dislocation-pair, Interstitial
			\\
		\end{tabular}
	\end{ruledtabular}
	\caption{Reference curvatures multipoles and possible realizations.}
	\label{tab:table1}
\end{table}

In  a continuum limit, the reference curvature describes distributed multipoles
\begin{eqnarray}
	\Kbar^0   = M(\xvec) + \nablabar_\alpha P^\alpha(\xvec) + \nablabar_{\alpha \beta} Q^{\alpha\beta}(\xvec) + \dots
	\label{eq:MultipoleExpansion}
\end{eqnarray} 
with $M$, $P$ and $Q$ distributions of disclinations, dislocations, and quadrupoles \cite{15MSK}. 
Singular multipoles are materialized via anelastic deformations which modify the reference metric. 
The simplest anelastic deformation is a local change in the reference state, 
\begin{equation}
	\gbar_{\alpha\beta} = \gbar^0_{\alpha\beta} + \deltan(\xvec)\, q_{\alpha\beta} \;.
\end{equation}
% \Noemie{should state $q$ is symmetric; I'm not sure about having $\gbar$ on both sides here}
The trace of $q$ corresponds to an area change, and the trace-less symmetric part corresponds to local shear. 
This type of metric deformation describes a wide variety of screening mechanism, as illustrated in \figref{fig:mechScreeningHierarchy}.
For small anelastic deformations the leading order of the reference curvature associated with $\gbar$ is
\begin{equation}
	\Kbar = \Kbar^0   + {Q}^{\alpha\beta} \nablabar_{\alpha\beta}  \delta(\xvec)
\end{equation}
with ${Q}^{\alpha\beta} = \bar\varepsilon^{\alpha\mu} \bar\varepsilon^{\beta\nu} q_{\mu\nu}$ and  $\bar\varepsilon$ are the Levi-Civita tensors with respect to $\gbar^0$ \cite{carroll2019spacetime}. {In light of the multipole expansion in \eqref{eq:MultipoleExpansion} we find that a local material rearrangement induces a localized quadrupolar elastic charge.}

This reflects a deeper property of elastic charges:    
In \cite{KMS2015} it was proved that the lowest order elastic multipole that can be nucleated by a local material deformation is quadrupolar. {The proof relies on global geometric properties which are impossible to change via local deformations. This geometric conservation law makes the elastic quadrupoles analogous to electric dipoles, which are the lowest order electric charges that can be nucleated locally without violating conservation of charge.} 
%\sout{This is analogous to electric dipoles in dielectric materials. }
The inevitable conclusion is that the quadrupolar field $Q^{\alpha\beta}(\xvec)$ is, in principle, the natural screening field in solids. 
Motivated by these observations we turn to derive a screening theory of elastic-like solids by accounting for induced quadrupoles and their nucleation cost. For that we briefly review the geometric approach to elasticity and the possible screening modes.

\subsection{Elasticity}
\label{sec:elasticity}
For a purely elastic material the reference metric $\gbar^0$ is fixed, and does not {change} in response to external loads. The elastic strain is then 
\begin{equation}
	\uel = \frac{1}{2} \left(g - \gbar^0\right) \Noemie{.}
\end{equation}
The equilibrium equations is derived from a mechanical free energy
\begin{equation}
	F =  \int_{\M} \left(\mathcal{U} - \mathcal{W}\right) \,\dif S_{\gbar^0}  - \int_{\partial \M}  \W_B\,\dif l_{\gbar^0}\;,
	\label{eq:FreeElEnergy}
\end{equation}
where $\U$ is the elastic energy density, while $\W$ and $\W_B$ encode the work density done on the system, e.g. by external forces acting either in the bulk or on the boundary, respectively. Upon assuming small strains, the elastic energy is Hookean
\begin{equation}
	\U   = \frac{1}{2} \A^{\alpha\beta\gamma\delta} \uel_{\alpha\beta}\uel_{\beta\gamma} \;.
    \label{eq:ElasticEnergy}
\end{equation}
In the absence of body forces and in the presence of traction forces the work densities are
\begin{eqnarray}
	\W  &=& 0 \\
	 \W_B &=& \mathbf{t} \cdot \mathbf{d}\;.
\end{eqnarray}
Here $\mathbf{d}$ is the displacement field defined relative to the ground-state, $\mathbf{t}$ are the imposed traction forces, and $\A$ is the elastic tensor encoding material properties. In a homogeneous and isotropic material
\begin{equation}
	\A^{\alpha\beta\gamma\delta} = \frac{\,\nu \,Y  }{1-\nu^2} \left(    \gbar^{\alpha\beta}\gbar^{\gamma\delta}+ \frac{1-\nu}{ 2\nu }(\gbar^{\alpha\gamma}\gbar^{\beta\delta} +   \gbar^{\alpha\delta}\gbar^{\beta\gamma}) \right)\,,
\end{equation}
with $Y$ the Young's modulus and $\nu$ Poissons' ratio. 
The stress tensor is defined by the variation of energy density with respect to the elastic strain, leading to Hooke's law
\begin{equation}
	\sigma^{\alpha\beta} = \A^{\alpha\beta\gamma\delta} u_{\gamma\delta}\Noemie{.}
\end{equation}
Upon minimizing \eqref{eq:FreeElEnergy} with respect to the embedding $\phi$ we obtain the equilibrium equation $\mathrm{div} \sigma = 0$, which takes the explicit form
\begin{equation}
	\bar{\nabla}_{\mu} \sigma^{\mu\nu} + \brk{\Gamma^\nu_{\alpha \beta}-\bar{\Gamma}^\nu_{\alpha \beta}} \sigma^{\alpha \beta} = 0,
	\label{eq:DivSigma}
\end{equation}
along with the boundary conditions
\begin{equation}
	n_{\alpha} \sigma^{\alpha \beta} = t^{\beta}\,.
	\label{eq:BC}
\end{equation}
This form of the equilibrium equation accounts for geometric nonlinearities and was first introduced in \cite{09ESK}, and is given in \appref{sec:EqEq}. A systematic method for solving it nonlinearly in the case of non-euclidean reference metric was introduced in \cite{14MSK}.

%%%%%%%%%%%%%%%%%%%%%%%%%%%%%%%%%%%%%%%%%%%%%%%%%%%%%%%%%%%%%%%%
\subsection{Screened Elasticity}
\label{sec:SE}
When strain relaxation mechanisms are available, the reference metric  is no longer fixed, but can evolve in response to deformations. We therefore distinguish between the (fixed) initial reference metric $\gbar^0$, and the temporary reference metric relative to which elastic deformations are measured
\begin{equation}
	\gbar = \gbar^0 + q\;.
 \label{eq:gbar}
\end{equation}
Here $q$ is the density of quadrupole perturbation to the reference metric $\gbar^0$. 
Correspondingly, the elastic tensor $\A$, covariant derivatives $\bar{\nabla}$, and the raising and lowering of indices are all defined with the fixed reference metric $\gbar^0$.
The elastic strain is the deviation of the current metric from the updated reference metric
\begin{eqnarray}
	\uel = \frac{1}{2} \left(g - \gbar \right) =  \frac{1}{2} \left(g - \gbar^0 - q \right) =  u - \frac{1}{2}q\;,
\end{eqnarray}
where $u = \tfrac{1}{2}(g - \gbar^0)$ is the total strain, measuring the deformation relative to the initial configuration.
The screened elastic energy stored in the system still  {has} the form \eqref{eq:ElasticEnergy}, 
\begin{equation}
	F_\mathrm{Sc} =  \int_{\M} \left(\mathcal{U} - \mathcal{W}\right) \,\dif S_{\gbar^0}  - \int_{\partial \M}  \W_B\,\dif l_{\gbar^0}\;,
	\label{eq:FreeElEnergy}
\end{equation}
with
\begin{eqnarray}
	\U &=&\frac{1}{2} \A^{\alpha\beta\gamma\delta} \uel_{\alpha\beta}\uel_{\beta\gamma}  = \frac{1}{2} \A^{\alpha\beta\gamma\delta} u_{\alpha\beta}u_{\beta\gamma}   \\&-&  \frac{1}{2}\A^{\alpha\beta\gamma\delta} u_{\alpha\beta}q_{\beta\gamma}  + \frac{1}{8} \A^{\alpha\beta\gamma\delta} q_{\alpha\beta}q_{\beta\gamma}  \notag \;.
	\label{eq:SCElasticEnergy}
\end{eqnarray}
This form of the energy uncovers the elastic interactions between the induced quadrupoles: 
the first term in the second row represents the elastic interaction between the quadrupole $q$ at point $\xvec$ with the background stress and all {the} other quadrupoles, and the last term represents the self-interaction elastic energy {corresponding to} the energy stored in the elastic field induced by a single quadrupole. Another important contribution to the self-interaction term is the work done on the system {in order to nucleate} the quadrupole core. This material dependent property is therefore contributing to the work term in \eqref{eq:FreeElEnergy} 
\begin{equation}
	\W = \W[q] \;.
\end{equation}
Here $\W$ is a functional whose specific form depends on the underlying screening mechanism and material properties. 

At this point we draw inspiration from the electrostatic analogue, specifically from \eqref{eq:Landau2} which builds on the multipole expansion, and write the general form of $\W$ reflecting screening by quadrupoles, dipoles, and monopoles
\begin{eqnarray}
	\W = \frac{1}{2} \Lambda^{\mathrm{Q}}_{\alpha\beta\gamma\delta} Q^{\alpha\beta}Q^{\gamma\delta}  +  \frac{1}{2} \Lambda^{\mathrm{P}}_{\alpha\beta} P^\alpha P^\beta + \frac{1}{2} \Lambda^{\mathrm{M}} M^2 \;, 
	\label{eq:cost}
\end{eqnarray}
where
\begin{eqnarray}
	Q^{\alpha\beta} = \bar\varepsilon^{\alpha\mu} \bar\varepsilon^{\beta\nu} q_{\mu\nu}\;, \, P^\alpha = \bar{\nabla}_\mu Q^{\alpha\mu}\; ,\,  M = \bar{\nabla}_{\alpha\beta} Q^{\alpha\beta} \;.
 \label{eq:qQ}
\end{eqnarray}
From homogeneity, isotropy, and the dimensions of $\W$ we find 
\begin{eqnarray}
     \Lambda^{\mathrm{Q}}_{\alpha\beta\gamma\delta} &=& \lambda_Q \gbar^0_{\alpha\beta} \gbar^0_{\gamma\delta} + \mu_Q \left(\gbar^0_{\alpha\gamma} \gbar^0_{\beta\delta} + \gbar^0_{\alpha\delta} \gbar^0_{\beta\gamma}\right) \notag \\
     \Lambda^{\mathrm{P}}_{\alpha\beta} &=& \tfrac{1}{2} Y \ell_P^2 \gbar^0_{\alpha\beta}\;, \notag \\ 
     \Lambda^{\mathrm{M}} &=& Y \ell_M^4 \;,     
     \label{eq:Moduli}
\end{eqnarray}
with $Y$ the Youngs modulus and $\ell_\mathrm{P},\ell_\mathrm{M}$  the typical length scales associated with each screening multipole.  
The quadrupole term in \eqref{eq:cost} represents the nucleation cost of a quadrupole field describing a distribution of local metric perturbations to $\gbar^0$. In this case the anelastic response of the material is quantified by the value of $Q$, describing the average uniform Eshelby-like deformation. This is similar to the weak screening by dislocation pairs (quadrupoles) in the solid phase of 2d crystalline materials. 
The second term in \eqref{eq:cost} describes the effective nucleation cost for dipoles that emerge from non uniform distribution of quadrupoles. 
In this case the anelastic response of the material is quantified by the spatial variation of $Q$ encoded in its divergence, and is similar to screening by dislocations (dipoles) in the hexatic phase of 2d crystalline materials. 
The last term in \eqref{eq:cost} describes the effective nucleation cost for monopoles, which is analogous to screening by disclinations (monopoles) in a melted 2d crystalline. 

\begin{figure}
    \centering
    \includegraphics[width = \columnwidth]{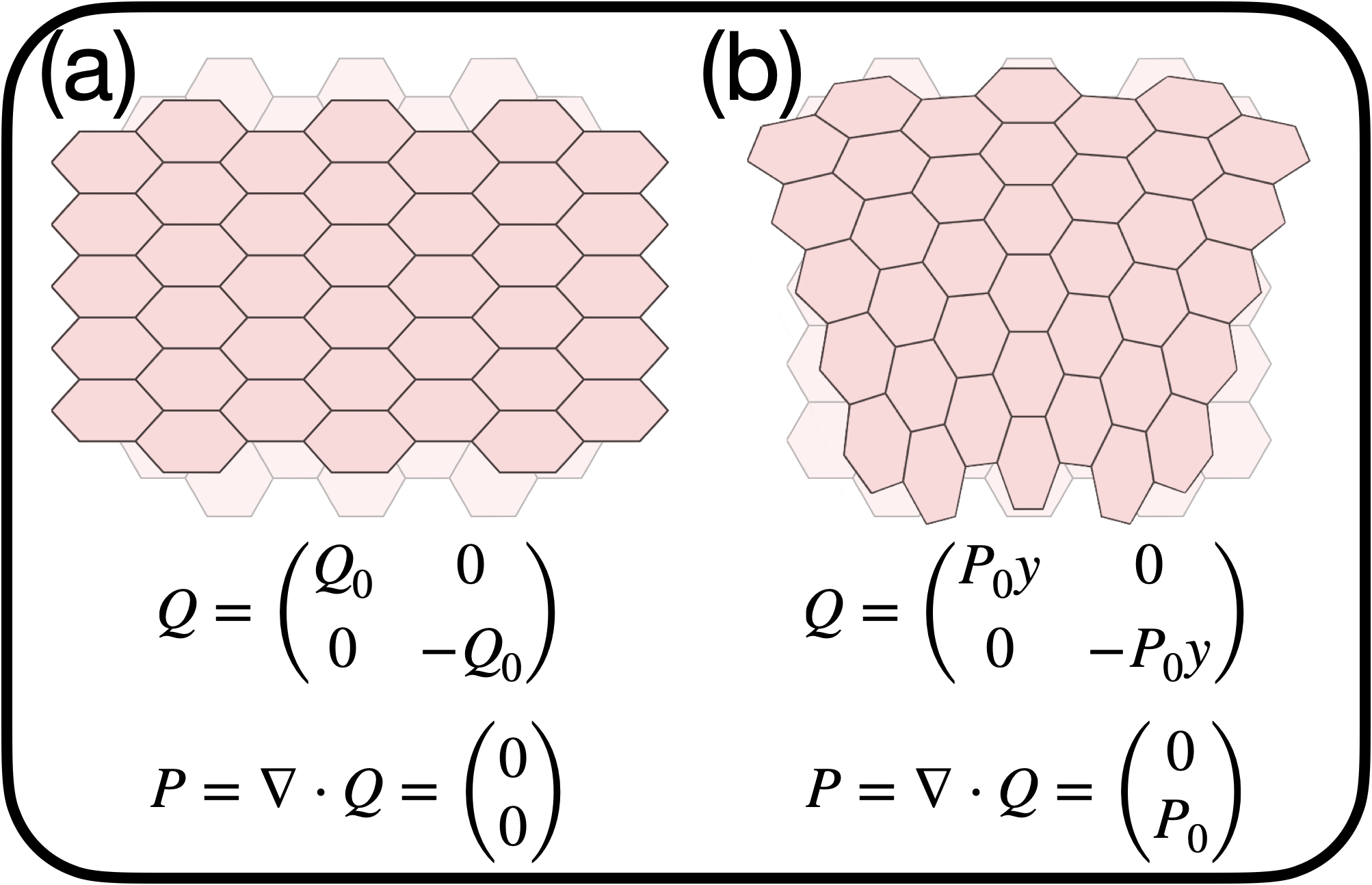}
    \caption{Anelastic deformations induced by (a) a uniform quadrupole and (b) a uniform dipole on a finite region. The deformed states are superimposed on the (semi-transparent) undeformed configuration.}
    \label{fig:ScreeningMultipole}
\end{figure}

The geometric realization of screening quadrupole and dipole is visualized in \figref{fig:ScreeningMultipole} where the semi-transparent and opaque configurations describe the rest states before and after the anelastic deformations, on a finite region. These anelastic deformations are derived by calculating the displacement field induced from uniform distribution of each multipole:
The deformation induced by a uniform $Q$ corresponds to a uniform strain and is visualized in \figref{fig:ScreeningMultipole}(a).
To interpret the dipole term we take a spatially varying quadrupole with uniform dipole $\mathbf{P} = P_0 \hat{y}$. The induced deformation is visualized in \figref{fig:ScreeningMultipole}(b), indicating  a non-Eshelby deformation that is of lower order in the multipole expansion. This is analogous to creating electric monopole from nonuniform dipole field.  
As for the monopole term in \eqref{eq:cost}, this screening mechanism induces non-zero curvature, thus  cannot be visualized via a planar deformation.

According to \eqref{eq:cost} and \eqref{eq:Moduli}, in principle all three screening mechanisms can act simultaneously. However, elastic materials corresponds to large $\ell_P$ and $\ell_M$, suppressing nucleation of dipoles and monopoles. When $\lambda_{\mathrm{Q}},\mu_{\mathrm{Q}} \to 0$ the nucleation cost of dipoles (the scale $\ell_P$) may become finite, and when $\ell_P \to 0$, the cost of monopoles ($\ell_M$) may become finite as well. 
This hierarchy of screening is based on scale separation of $\ellp, \ell_M$ and is in line with {the} scale-separation discussed after \eqref{eq:Landau2} in the electrostatic analog. 
It is also analogous to the hierarchy of Solid-Hexatic-Liquid phases, where dipole and monopole screenings correspond to the unbinding of dislocations (dipoles) and disclinations (monopoles) with finite nucleation energy in the hexatic and liquid phases, respectively \cite{nelson1979dislocation,nelson2002defects}.The mapping between the theories is discussed in Sec.~\ref{sec:Sum}. 

In light of this argument, in what follows we study the mechanics of the three screening modes separately, and we assume three {distinct} situations in which each {of the} terms in \eqref{eq:cost} dominates.

\section{Equilibrium equations}
\label{sec:Equilibrium}
Here we derive equilibrium equations for each of the quadrupole, dipole, and monopole screening regimes. The detailed derivation is given in \appref{sec:EqEq}.

The equilibrium equations are derived using the variation of an energy with respect to the embedding $\phi$ describing the configuration, and the induced quadrupole field $q$. 
Since $\W$ is independent of the configuration, the variation with respect to $\phi$ is the same in the different screening regimes. 
Explicitly, the mechanical free energy to be minimized is
\begin{eqnarray}
	F & = & \int_{\M} \left( \frac{1}{2} \A^{\alpha\beta\gamma\delta} \uel_{\alpha\beta}\uel_{\gamma\delta}  - \W[q] \right) \,\dif S_{\gbar^0} \notag\\
	& - & \int_{\partial \M}  \mathbf{t} \cdot \mathbf{d} \,\dif l_{\gbar^0}\;.
	\label{eq:FreeEnergyQuad}
\end{eqnarray}
Upon defining the elastic stress
\begin{eqnarray}
	\sigmael^{\alpha\beta} = \A^{\alpha\beta\gamma\delta} \uel_{\gamma\delta} =\frac{1}{2} \A^{\alpha\beta\gamma\delta} \left(g_{\gamma\delta} - g^0_{\gamma\delta} - q_{\gamma\delta} \right)
	\label{eq:ScreenedConsRel}
\end{eqnarray}
we find
the equilibrium equation
\begin{eqnarray}
	\bar{\nabla}_{\mu} \sigmael^{\mu\nu} + \brk{\Gamma^\nu_{\alpha \beta}-\bar{\Gamma}^\nu_{\alpha \beta}} \sigmael^{\alpha \beta} &=& 0 
	 \label{eq:solEQ}
\end{eqnarray}
along with the boundary conditions
\begin{equation}
	n_{\alpha} \sigmael^{\alpha \beta} = t^{\beta}\,
	\label{eq:BCquad}
\end{equation}
justifying our definition of the elastic stress tensor. 
We emphasize that from the solutions for the stress $\sigmael$ and the induced charges $q$ we can recover the actual metric through
\begin{equation}
	g_{\alpha\beta} = \gbar^0_{\alpha\beta} + q_{\alpha\beta} + 2 \A_{\alpha\beta\gamma\delta} \sigmael^{\gamma\delta}  \;.
\end{equation}
Here the notation $\A_{\alpha\beta\gamma\delta}$ is the inverse elastic tensor.

To recover the actual metric and configuration in equilibrium, \eqref{eq:solEQ} should be supplemented with an equation for the induced screening charges, obtained by {varying the energy \eqref{eq:FreeEnergyQuad}} with respect to $q$
\begin{eqnarray}
	\delta_q F & = & \int_{\M} \left(-\frac{1}{2} \sigmael^{\alpha\beta} \delta q_{\alpha\beta}  - \delta_q \W\right)  \,\dif S_{\gbar^0}
	\;.
	\label{eq:VarFq}
\end{eqnarray}
Next we perform the variation of $\W$, which is shown to strongly depends on the specific  screening regime.

\noindent\textbf{\textit{Quadrupole screening:}} In this case 
\begin{equation}
\begin{split}
        \W &= \frac{1}{2} \Lambda^{\mathrm{Q}}_{\alpha\beta\gamma\delta} Q^{\alpha\beta}Q^{\gamma\delta} = \frac{1}{2} \Lambda_{\mathrm{q}}^{\alpha\beta\gamma\delta} q_{\alpha\beta}q_{\gamma\delta}\;,
\end{split}
\label{eq:QuadW}
\end{equation} 
with $ \Lambda_{\mathrm{q}}$ proportional to $ \Lambda^{\mathrm{Q}}$ (see \appref{app:RenormalizedA}). 
Upon varying the the total energy with respect to $q$ we find
a linear relation between the induced quadrupole and the elastic stress
\begin{eqnarray}
	\sigmael^{\alpha\beta} + 2 {\bar{\varepsilon}}^{\alpha\mu}{\bar{\varepsilon}}^{\beta\nu} {\Lambda}^{\mathrm{Q}}_{\mu\nu\gamma\delta} Q^{\gamma\delta} &=& 0\;.
	\label{eq:solquad}
\end{eqnarray}
In analogy to models for dielectric media, such as the Maxwell-Garnett model \cite{choy2015effective,garnett1904xii}, this screening regime describes a material containing a dilute distribution of quadrupoles induced in response to external loads.

At this point we can integrate out the quadrupolar degree of freedom by substituting $q$ either in the constitutive relation Eq.~\ref{eq:ScreenedConsRel} or the energy in Eq.~\ref{eq:FreeEnergyQuad}. In both cases we end up with an effective elastic energy $F_Q$ that {only} depends on the total strain
\begin{eqnarray}
	F_Q &=& \int_{\M}  \frac{1}{2} \tilde{\A}^{\alpha\beta\gamma\delta} u_{\alpha\beta}u_{\gamma\delta} \,\dif S_{\gbar^0} 
	- \int_{\partial \M}  \mathbf{t} \cdot \mathbf{d} \,\dif l_{\gbar^0}\;,
\end{eqnarray}
{where $\tilde{\A}$ is}  an effective elastic tensor given {explicitly} in \appref{app:RenormalizedA},
encoding the mechanical effect of the induced quadrupoles, leading to a quasi-elastic theory.

{This result is also similar to dielectrics, where screening by electric dipoles re-scales the dielectric constants without otherwise modifying the theory.}

\noindent\textbf{\textit{Dipole screening:}}
In this case 
\begin{equation}
    \W = \frac{1}{2} \Lambda^{\mathrm{P}}_{\alpha\beta} P^\alpha P^\beta =\frac{1}{2} \Lambda^{\mathrm{P}}_{\alpha\beta} (\bar{\nabla}_\mu Q^{\alpha\mu}) ( \bar{\nabla}_\nu Q^{\beta\nu})  \;.
\end{equation}
Upon substituting the relation between $Q$ and $q$, and varying $\W$ with respect to $q$ we find
\begin{eqnarray}
	\sigmael^{\alpha\beta} +  \tfrac{1}{2}Y \ell_P^2 {\bar{\varepsilon}}^{\mu\alpha}{\bar{\varepsilon}}^{\nu\beta} \left( \bar\nabla_\mu P_\nu + \bar\nabla_\nu P_\mu\right) = 0
 \label{eq:DipoleEquilibrium}
\end{eqnarray}
along with the boundary condition
\begin{eqnarray}
	{\bar{\varepsilon}}^{\mu\alpha}{\bar{\varepsilon}}^{\nu\beta} \left( n_\mu P_\nu + n_\nu P_\mu\right) = 0\;.
	\label{eq:DipoleBC}
\end{eqnarray}

 Contrary to the quadrupole screening regime where a linear relation between stress and induced quadrupoles holds, here the stress is linearly proportional to the second gradient of the induced quadrupole field. An immediate consequence is the relation between elastic pressure and the induced isotropic quadrupole
\begin{eqnarray}
	\mathrm{Tr}\, \sigmael = \gbar_{\alpha\beta} \sigmael^{\alpha\beta} = - Y \ellp^2 \bar{\nabla}_{\mu\nu} Q^{\mu\nu}
	\label{eq:TrDip}
\end{eqnarray}
This situation is similar to its electrostatic analog, wherein a dielectric the induced dipoles are linearly proportional to the electric field, whereas in Debye-Hückel theory the electric field is proportional to the second gradient of the induced dipoles, as in \eqref{eq:Debye}.

\noindent\textbf{\textit{Monopole screening:}}
In this case 
\begin{equation}
    \W = \frac{1}{2} \Lambda^{\mathrm{M}} M^2= \frac{1}{2} \Lambda^{\mathrm{M}} (\bar{\nabla}_{\alpha\beta} Q^{\alpha\beta}) ( \bar{\nabla}_{\gamma\delta} Q^{\gamma\delta})
\end{equation}
and from the variation of $\W$ we find

\begin{eqnarray}
	\sigmael^{\rho\sigma} + Y \ell_M^4 \varepsilon^{\gamma\rho}\varepsilon^{\delta\sigma} (\bar{\nabla}_{\gamma\delta }\bar{\nabla}_{\alpha\beta} Q^{\alpha\beta}) =0
	\; \Noemie{\sout{.}}
	\label{eq:QuadMon}
\end{eqnarray}
with the boundary condition
\begin{eqnarray}
	{\bar{\varepsilon}}^{\mu\alpha}{\bar{\varepsilon}}^{\nu\beta} \left( n_\mu \bar\nabla_{\nu}M +n_\nu \bar\nabla_{\mu}M\right) = 0\;.
	\label{eq:MonopoleBC}
\end{eqnarray}

As in the dipole screening {regime}, here {too} we will find that the pressure, that is the trace of stress, is useful when integrating out the quadrupolar degree of freedom, and it takes the form
\begin{eqnarray}
	\mathrm{Tr} \, \sigmael =- \Lambda^{\mathrm{M}} (\bar{\Delta} \bar{\nabla}_{\alpha\beta} Q^{\alpha\beta}) 
	\; .
	\label{eq:TrMon}
\end{eqnarray}

In summary, the equilibrium equations for each screening mode are
\begin{equation}
    \begin{aligned}
        &\text{Equation}  &&\text{Mode} \\
        &\sigmael^{\alpha\beta} =-{\bar{\varepsilon}}^{\alpha\mu}{\bar{\varepsilon}}^{\beta\nu} {\Lambda}^{\mathrm{Q}}_{\mu\nu\gamma\delta} Q^{\gamma\delta}  \quad &&\text{Quadrupole} \\
        &\sigmael^{\alpha\beta} =-  \tfrac{1}{2}Y \ell_P^2 {\bar{\varepsilon}}^{\mu\alpha}{\bar{\varepsilon}}^{\nu\beta} \left( \bar\nabla_\mu P_\nu + \bar\nabla_\nu P_\mu\right)  \quad &&\text{Dipole} \\
                &\sigmael^{\alpha\beta} =- Y \ell_M^4 \bar\varepsilon^{\gamma\alpha}\bar\varepsilon^{\delta\beta} (\bar{\nabla}_{\gamma\delta }\bar{\nabla}_{\mu\nu} Q^{\mu\nu}) \quad &&\text{Monopole}
    \end{aligned}
	\label{eq:ScreeningEq}
\end{equation}

\section{Potential Theory} 
\label{sec:Potential}
To solve the equilibrium equations for the stress and the induced charges, we develop a potential theory generalizing the Airy stress function approach. In this approach a representation of the stress solving \eqref{eq:solEQ} is given in terms of a scalar function
\begin{equation}
	\sigmael^{\mu \nu} = \frac{1}{\sqrt{|\gbar|}}   \frac{1}{\sqrt{|g|}} {\varepsilon}^{\mu \alpha}  \varepsilon^{\nu \beta} \nabla^g_{\alpha\beta} \chi
	\label{eq:HexSolRep}
\end{equation}

A geometric compatibility condition is needed to determine the stress function $\chi$, that is requiring the Gaussian curvature of the actual metric $g$ to vanish. 
From the definition of stress and strain we get {an expression for the actual metric}
\begin{eqnarray}
	g_{\alpha\beta} = \gbar_{\alpha\beta}^0 + \varepsilon_{\alpha\mu}\varepsilon_{\beta\nu} Q^{\mu\nu} + 2 \A_{\alpha\beta\gamma\delta} \sigmael^{\gamma\delta} {,}
\end{eqnarray}
{which} is implicit {due to} the complicated dependence of $\sigmael$ on $g$. 

To calculate the curvature of $g$ and enforce the geometric compatibility condition we now assume that both {the} elastic and {the} total strains are small, that is $g \approx \gbar \approx \gbar^0$. Within this approximation a perturbative expansion for the stress potential is applicable. The leading order {term of the} curvature takes the form
\begin{equation}
	0 = \bar{K}^0 + \nabla_{\alpha\beta}Q^{\alpha\beta} - \frac{1}{Y} \Delta\Delta \chi
	\label{eq:tmpBiHex}
\end{equation} 
The term $\nabla_{\alpha\beta}Q^{\alpha\beta}$ represent the induced effective monopoles, which depends on the specific screening regime.
To close the equation, and integrate out the quadrupolar degrees of freedom, we determined the induced effective monopoles by substituting \eqref{eq:HexSolRep} in Eq.~\ref{eq:ScreeningEq}.
We find (see \appref{app:InducedMonopoles} for details)
\begin{equation}
    \nablabar_{\alpha\beta} Q^{\alpha\beta} = -\frac{1}{Y} \times
    \left\{
    \begin{array}{ll} 
      0 & \text{Quadrupole} \\
    \ellp^{-2}\bar{\Delta} \chi & \text{Dipole} \\
      \ell_M^{-4}  \chi & \text{Monopole}
    \end{array}
    \right.
    \label{eq:IndMon}
\end{equation}
In the third equation, corresponding to the monopole screening regime, the induced monopole is determined up to an arbitrary function satisfying $\nablabar_{\alpha\beta}\chi_g = 0$, and we choose a gauge with $\chi_g=0$.

{Having found} the explicit expression of $\bar{\nabla}_{\alpha\beta} Q^{\alpha\beta}$ in each screening mode, equation \eqref{eq:tmpBiHex} is now closed
\begin{equation}
    \begin{aligned}
        &\text{Screened Stress Function}  \quad &&\text{Mode} \\
        &\frac{1}{\tilde Y}\Delta \Delta \chi  = \bar{K}^0  \quad &&\text{Quadrupole} \\
        &\frac{1}{\tilde Y}\Delta \Delta \chi + \frac{1}{ \tilde Y} \ellp^{-2} \Delta \chi  = \bar{K}^0  \quad &&\text{Dipole} \\
                &\frac{1}{ \tilde Y}\Delta \Delta \chi + \frac{1}{\tilde Y} \ell_M^{-4} \chi  = \bar{K}^0  \quad &&\text{Monopole}
    \end{aligned}
    \label{eq:ScreenedSF}
\end{equation}
These equations derived based on the assumption of scale separation, discussed in the introduction. Within this assumption 
we can combine them into one equations that holds when screening is dominated by either quadrupole, dipole, or monopole charges
\begin{equation}
\Delta \Delta \chi +  \ell_P^{-2} \Delta \chi + \ell_M^{-4}  \chi  =  Y \bar{K}^0\;.
\label{eq:unified}
\end{equation}
Once the equation for $\chi$ is solved the stress tensor can be calculated and boundary conditions enforced to uniquely determine $\chi$. However, to recover the displacement field it is required to calculate the actual metric of the embedding, and therefore the induced quadrupoles. For that the solution for the elastic stress is substituted in \eqref{eq:ScreeningEq} which then should be solved for the induced quadrupoles, subjected to the boundary conditions (\eqref{eq:DipoleBC} in the dipole regime and \eqref{eq:MonopoleBC} in the monopole regime).

At this point we identify an explicit link with the theory of melting in 2d crystals. It was recently shown that the theory of defects-mediated melting is dual to a sine-Gordon like hamiltonian \cite{zhai2019two,pretko2019crystal}. Upon deriving the equilibrium equations from the proposed hamiltonian the equation in \eqref{eq:unified} are recovered.
This observation suggests that the dipole screening regime developed in this work forms a mechanical realization of the hexatic phase, that is traditionally associated with structural properties.

A comment on gauge freedom is necessary at this point: One may suspect that the explicit dependence of \eqref{eq:unified} on the value of the stress function $\chi$ violates the gauge freedom of the stress tensor. However, this only reflects the gauge choice made when solving for the induced effective monopole in the monopole screening regime \eqref{eq:IndMon}.
This is similar to loss of gauge freedom in Debye-Hückel theory as in \eqref{eq:Helmholtz}. 

% Upon substituting the representation of the elastic stress in terms of the stress-function the solutions follow immediately and read
% \begin{equation}
%     \begin{aligned}
%         & Q^{\alpha\beta} = -\frac{1}{2} {\Lambda}_{\mathrm{Q}}^{\alpha\beta\gamma\delta} \nablabar_{\alpha\beta} \chi  \quad &&\text{Quadrupole} \\
%         &P_\mu =P^0_\mu -  \ell_P^{2}  \nablabar_{\mu}\chi  &&\text{Dipole} \\
%                 &M =- \ell_M^{4}\left(\chi - \chi_G\right)  &&\text{Monopole}
%     \end{aligned}
% 	\label{eq:ScreeningBC}
% \end{equation}
% where $P^0$ is a parallel vector field and $\chi_G$ is a scalar function satisfying the gauge freedom $\bar{\varepsilon}^{\alpha\mu}\bar{\varepsilon}^{\beta\nu} \bar{\nabla}_{\mu\nu} \chi_G = 0$.
% These solutions are subjected to the boundary conditions calculated before in \eqref{eq:DipoleBC} and \eqref{eq:MonopoleBC}, and can now be expressed in terms of the stress function 
% \begin{equation}
%     \begin{aligned}
%         & \text{------}  \quad &&\text{Quadrupole} \\
%         & \nablabar_{\mu}\chi = P^0_\mu/\ell_P^{2} &&\text{Dipole} \\
%                 &\chi = \chi_G - M_0/\ell_M^{4}   &&\text{Monopole}
%     \end{aligned}
% 	\label{eq:ScreeningSFEq}
% \end{equation}

\section{Applications}
\label{sec:Applications}
The hierarchical form of \eqref{eq:cost} suggests that solids with quadrupolar relaxation mechanism are prone to dipole screening. This hypothesis, if true, unifies a variety of systems that are fundamentally different from each other, under the same screening theory.
For example, cellular epithelial tissue respond to mechanical loads by cell rearrangements \cite{Bi2015rigidity, park2015unjamming} and shape changes\cite{moshe2018geometric,hernandez2022anomalous}, both quadrupolar in nature. 
Holes in perforated (``holey'') mechanical meta-materials release stresses by forming imaginary quadrupoles \cite{matsumoto2009elastic, bar2020geometric}. Non-uniform hole sizes, as in disordered metamaterials, will induce spatially varying quadrupoles, potentially leading to dipole screening.
Last but not the least, screening can form in wrinkled and crumpled thin sheets. The system shown in \figref{fig:mechScreeningHierarchy}(d) demonstrate the quadrupolar nature of local wrinkles, that can merge to form long wrinkles, as observed in other scenarios such as \cite{hure2012stamping}. If a wrinkle ends at the bulk it leaves a free dipole, supporting the possibility of dipole screening.
We therefore expect our theory to form an effective 2d description of certain wrinkled systems, holey metamaterials, glasses, tissue models and granular matter.

In the next subsections we study the mechanical implications of dipole and monopole screening on prototypical mechanical scenarios such as the  fields induced by sources of stresses (defects), and the interactions between them.

\subsection{Screened Green's function}
A prominent  manifestation of screening is the modified form of the potential associated with a point monopole charge. 
This potential is of importance for two main reasons: (i) Its functional form characterizes the nature and effect of screening, and (ii) It forms a Green's function for the non-homogeneous equation \eqref{eq:ScreenedSF}.
 \begin{figure}
    \centering
    \includegraphics[width = 0.9\columnwidth]{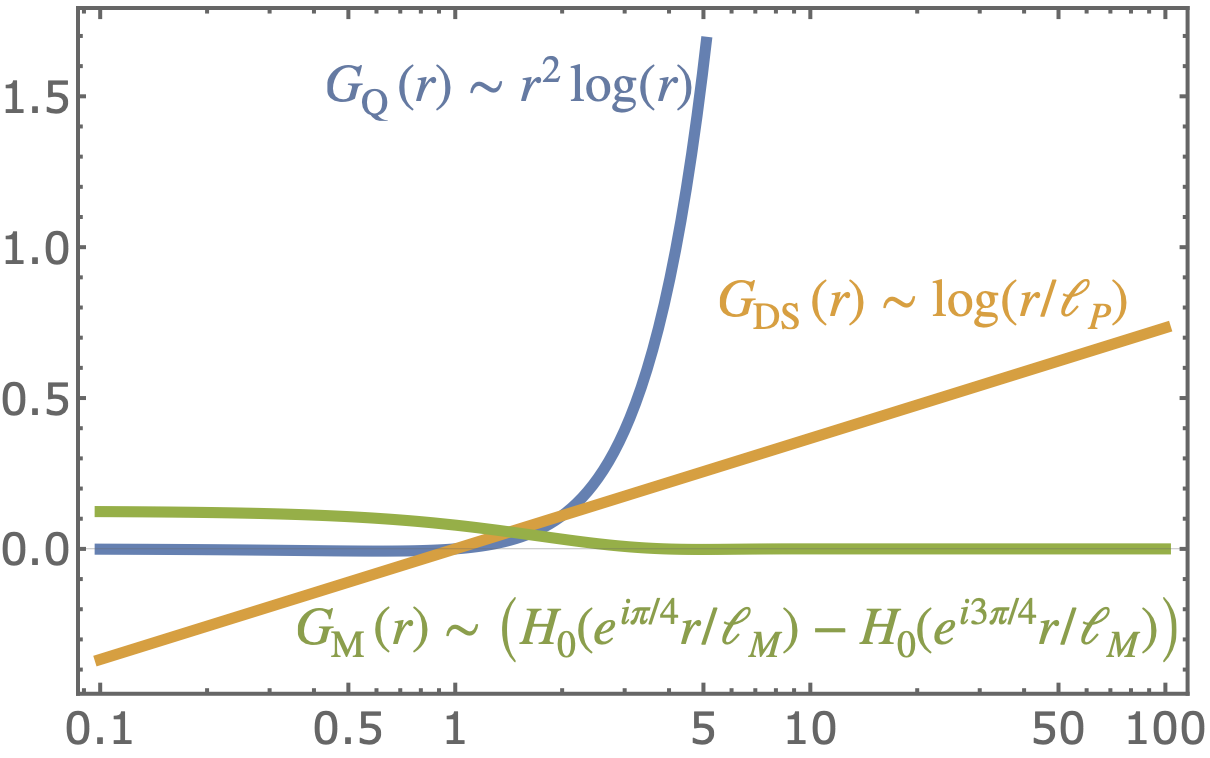}
    \caption{Green's functions associated with the inhomogeneous screened equations \eqref{eq:ScreenedSF} plotted on a semi-log scale. The blue, yellow and green curves represent the stress function associated with a monopole screened by quadrupoles, dipoles, and monopoles.}
    \label{fig:Greens}
\end{figure}

Monopolar elastic charges can be created by removal or insertion of an angular section.
In hexagonal crystalline structures they form 5- or 7-fold disclinations. A metric description of defects generalizes the concept of structural defects to solids with no underlying order, e.g. amorphous solids \cite{moshe2015geometry, KMS2015}. 
In analogy with the screened fundamental-solution in Debye-Hückel theory, known as Yukawa potential, we solve \eqref{eq:ScreenedSF} in each screening regime for a monopolar source term $\Kbar^0 = \delta (\xvec)$. 

To solve the equations it is useful to define a Helmholtz operator
\begin{equation}
    \mathcal{H}^{\theta}_{\ell} = \Delta + e^{\mathrm{i} \theta} \ell^{-2}\;,
\end{equation}
with which \eqref{eq:ScreenedSF} reads
\begin{equation}
    \begin{aligned}
        &\frac{1}{\tilde Y}\mathcal{H}^0_{0} \mathcal{H}^0_{0} \chi  = \bar{K}^0  \quad &&\text{Quadrupole} \\
        &\frac{1}{\tilde Y} \mathcal{H}^0_{\ellp} \mathcal{H}^0_{0} \chi  = \bar{K}^0  \quad &&\text{Dipole} \\
                &\frac{1}{ \tilde Y} \mathcal{H}^{\pi/4}_{\ell_M}
    \mathcal{H}^{-\pi/4}_{\ell_M} \chi   = \bar{K}^0  \quad &&\text{Monopole}
    \end{aligned}
\end{equation}

An important property of $\mathcal{H}$ is that the kernels of two different operators are disjoint. Therefore the homogeneous equation in the case of dipole and monopole screenings reduce to pairs of second order equations.

In the case of quadrupole screening the Green's function $G^{QS}$ coincides with the classical solution of a single disclination. 
To find the solution in the case of dipole screening we write the general polar symmetric solutions of the two equation $\mathcal{H}^0_{0} \chi_D  = 0$ and $\mathcal{H}^0_{\ellp} \chi_D  = 0$, hence
\begin{eqnarray}
     \chi^{DS}(r) = c_1 \log(r/\ellp) + c_2 J_0(r/\ellp) + c_3 Y_0(r/\ellp) + c_4\;,
     \label{eq:GreensDipole}
\end{eqnarray}
Similarly, the solution in the case of monopole screening is found by solving $\mathcal{H}^{-\pi/4}_{\ell_M} \chi_M  = 0$ and $\mathcal{H}^{\pi/4}_{\ell_M} \chi_M  = 0$, and reads
\begin{eqnarray}
    \chi^{MS}(r) &=& d_1 J_0 \left(e^{\tfrac{\pi \mathrm{i}}{4} }\, {r }/{\ell_M} \right) + d_2 J_0 \left(e^{\tfrac{3 \pi \mathrm{i}}{4} }\, {r }/{\ell_M} \right) \notag \\ 
    &+&d_3 Y_0 \left(e^{\tfrac{\pi \mathrm{i}}{4} }\, {r }/{\ell_M} \right) + d_4 Y_0 \left(e^{\tfrac{3 \pi \mathrm{i}}{4} }\, {r }/{\ell_M} \right) \notag 
\end{eqnarray}
The coefficients $c_i$ and $d_i$ are determined by boundary conditions, and by a topological condition obtained by integrating both sides of \eqref{eq:ScreenedSF} with $\Kbar = \delta(\xvec)$ over the area. In the case of monopole screening we also set the value of stress-function at infinity, reflecting the gauge choice made in \eqref{eq:IndMon}. 
The case of traction-free boundary conditions in a finite systems is detailed in \appref{app:2}.
Green's function is obtained by solving the problem in an infinite system with vanishing stress at infinity. The solutions for the three screening regimes  are plotted in \figref{fig:Greens} and are given by
\begin{equation}
\begin{split}
    G^\text{QS}(\xvec,\xvec') &= \frac{Y\, |\xvec - \xvec'|^2}{8\pi}  \log\frac{|\xvec - \xvec'|}{\ellp}\;,  \\
     G^\text{DS}(\xvec,\xvec') &= \frac{Y\,\ellp^2}{2\pi}  \log\frac{|\xvec - \xvec'|}{\ellp}\;,  \\
     G^\text{MS}(\xvec,\xvec') &=   \\
     \frac{Y \ell_M^2 }{8} & \left[ H_0\left({e^{\frac{i \pi }{4}}}  \frac{|\xvec - \xvec'|}{\ell_M} \right) -H_0\left({e^{-\frac{i \pi }{4}}}  \frac{|\xvec - \xvec'|}{\ell_M} \right) \right]
\end{split}     
\label{eq:DipScGreen}
\end{equation}
with  $H_0$ the Hankel function defined by
\begin{eqnarray}
     H_0(z) &=& J_0 (z) + \mathrm{i} Y_0 (z) \;.
\end{eqnarray}

% \begin{eqnarray}
%     \chi(r) &=& c_1 J_0 \left(\frac{r \, e^{\tfrac{\pi \mathrm{i}}{4} }}{\ell_M} \right) + c_2 J_0 \left(\frac{r \, e^{\tfrac{3\pi \mathrm{i}}{4} }}{\ell_M} \right) \notag \\ 
%     &+& c_3 Y_0 \left(\frac{r \, e^{\tfrac{\pi \mathrm{i}}{4} }}{\ell_M} \right) + c_4 Y    _0 \left(\frac{r \, e^{\tfrac{3\pi \mathrm{i}}{4} }}{\ell_M} \right)  \;.
% \end{eqnarray}

The Green's function screened by dipoles $G^\text{DS}$ in \eqref{eq:DipScGreen} is consistent with the potential induced by a disclination in the hexatic phase, and forms the basis for the sequential transition from hexatic to fluid phase.
Furthermore, this result provides a potential explanation for a problem presented in a visionary study Ref. \cite{chaudhari1979edge}. In that work the authors studied the elastic fields induced by edge- and screw- dislocations in a Lennard-Jones model of amorphous solid. They discovered that the stress fields of a screw-dislocation are elastic-like, whereas those of edge-dislocation are smeared out. 
In our theory an edge dislocation is dipolar and therefore is significantly screened by dipoles as expressed by $G^\text{DS}$. This is contrary to the screw dislocation which is not dipolar and therefore cannot be effectively screened by dipoles. A systematic study of this problem from the perspective of screening that will compare theoretical predictions with numerical simulations of amorphous solids is left for a future work.

\subsection{Screened geometric charges and their interactions}
In this section we highlight several key results that follow from the fundamental solution $G^{\mathrm{DS}}$ in the dipole-screening regime. Additionally, we study the interactions between screened geometric charges. .

It was shown in \cite{moshe2015geometry, KMS2015} that defects and other sources of stresses can be defined geometrically regardless of a specific physical model. In this theory sources of stresses are singularities of $\Kbar$. For example, dislocation correspond to $\Kbar = \mathbf{b} \cdot \nabla \delta (\xvec)$, and isotropic Eshelby inclusion corresponds to $\Kbar = p \Delta \delta (\xvec)$. 

From linearity of \eqref{eq:ScreenedSF}, and from commutation of derivatives with the $\mathcal{H}$ operator, taking the derivative of both sides with $\Kbar = \delta(\xvec)$ yields new solutions for higher order sources of stresses. 

For example, the stress function of a dipole described by $\Kbar = \mathbf{b} \cdot \nabla \delta (\xvec)$ and
analogous to a dislocation in the hexatic phase, is
\begin{eqnarray}
    \chi_\mathrm{\mathbf{b}} = \mathbf{b} \cdot \nabla G^\mathrm{DS} = \frac{Y \ellp^2}{2\pi} \frac{\mathbf{b}\cdot \xvec}{ r^2}\;.
\end{eqnarray}
The stresses derived from this solution decay rapidly with $r$. Upon substituting in the energy density one finds that the total energy of a screened dislocation converges at infinite systems, and reflects only the core energy. 

The second example is that of an isotropic Eshelby inclusion, whose solution is
\begin{eqnarray}
    \chi_\mathrm{Iso}(r) = p \Delta G^\mathrm{DS}(r) = p \delta(\xvec)\;.
\end{eqnarray}
This indicates that an isotropic inclusion in an infinite medium will be completely screened by emergent dipoles. It is important to note that the response to a localized expansion in a finite system is different (see solution in \appref{app:2}), and it exhibits spatial oscillations as previously reported by some of the authors \cite{SchreiberKeren2021, lemaitre2021anomalous,mondal2022experimental,kumar2022anomalous, bhowmik2022direct,charan2022Anomalous}.

The stress-functions of the screened dislocation and isotropic inclusion solve the homogeneous equation \eqref{eq:ScreenedSF}, thus in the kernel of the relevant differential operator. A comprehensive analysis of the kernel of $\mathcal{H}^\varphi_{\ell}$ is needed in order to classify and derive all singular solutions, and is an ongoing research topic and will be pursued in future studies.
 
% \subsection{Interaction between geometric charges}
% In this section 
Next we examine the interactions between screened sources of stress. It is well established that in the elastic regime, the energy stored in the medium can be represented by the stress function and charge distribution \cite{moshe2015elastic}
\begin{equation}
    U = \int \chi \,\Kbar \, \mathrm{dVol} \;.
    \label{eq:Interaction}
\end{equation}
In \appref{app:Int} we show that this relation holds also in the screened regime, hence we can use it to study the interactions between basic sources of stresses. 
For example, it is known that isotropic inclusions do not interact in the elastic framework \cite{moshe2015elastic}. However, still in the elastic framework, a disclination do interact with an inclusion. This is seen by taking $\Kbar_\mathrm{disc} = q \delta (\xvec)$ and $\chi_\mathrm{Iso} = p \log (\xvec - \xvec_0)$. 
From \eqref{eq:Interaction} we find that the interaction between an inclusion and disclination is
\begin{equation}
    U = q \, p \log(r) \;.
    \label{eq:Interaction}
\end{equation}
where $r$ is the distance between the two charges. 

In the case of dipole screening we still have $\Kbar_\mathrm{disc} = q \delta (\xvec)$, however, the screened stress function of the inclusion is $\chi^{DS}_\mathrm{Iso} = p \delta (\xvec - \xvec_0)$. In that case the interaction is zero, and the induced dipole field completely screen out the interaction. The interactions between other multipoles is calculated in the same way.

\section{Summary and discussion}
\label{sec:Sum}
In this work, we developed a hierarchy of continuum screening theories that generalize classical elasticity and are expected to be applicable to a variety of different solid-like systems, such as granular materials, cellular tissue, and mechanical metamaterials. While the traditional approach to non-mechanical screening theories is based on statistical and thermodynamic arguments, our theory is based on geometric arguments under the assumption that a long-wavelength description of screened solids is valid.

Based on the conservation laws associated with the geometry of two-dimensional Riemannian manifolds, our theory predicts three states of solid-like matter: Quasi-elastic quadrupole-screened, anomalous dipole-screened, and monopole-screened solids.
The case of dipole screening exhibits mechanical behavior that is similar to the hexatic phase, and thus forms an intermediate state between a solid and liquid. The existence of dipole screening has been fully confirmed in a series of recent works on granular and glassy matter. The predictions from the monopole screening regime have not yet been observed in athermal systems.

Our findings suggest that the current understanding of the jamming transition in granular matter is incomplete. For example, it is widely accepted that upon decreasing the pressure from a dense granular material, at a critical packing fraction, the material undergoes an unjamming transition to a liquid-like state that does not support shear.
Instead, based on our theory, we expect a sequential transition from a dense granular solid, to a dipole-screened solid-like state, and then to an unjammed state described by monopole screening, similar to the liquid state in the melting of two-dimensional crystals.

The effect of mechanical screening, in principle, is not limited to quasi-static deformations, as studied in this work, and is expected to have implications on the mechanics of both inertial and dissipative systems. Furthermore, well-studied phenomena such as fracture can now be studied within the framework of screened elasticity. These and other research questions are left for future study.

\acknowledgments
We would like to thank Mokhtar Adda-Bedia, Leo Radzihovsky and Keren Schreiber-Re'em for stimulating discussions.  The research was supported by the Israel Science Foundation grant No. 1441/19.

%%%%%%%%%%%%%%%%%%%%%%%%%%%%%%%%%%%%%%%%%%%%%%%%%%%%%%%%%%

\appendix

%%%%%%%%%%%%%%%%%%%%%%%%%%%%%%%%%%
\section{Derivation of Equilibrium Equation for the elastic stress}
\label{sec:EqEq}
In this section we derive the nonlinear equilibrium equations for the elastic stress and the corresponding boundary conditions.  

An important quantity that will come back later is the coordinate transformation of a vector from one coordinate system to another. Consider two manifolds $\M,\N$ on which coordinate systems are denoted with Greek indices $\mu,\nu,...$, and roman indices $i,j,..$ respectively. Given a mapping $\phi:\M \to \N$ the transformation of a vector from $\M$ to $\N$ is given by
\begin{eqnarray}
    v_{\N}^i = \frac{\partial \phi^i }{\partial x^\mu} v_\M^\mu\;.
\end{eqnarray}
 
The material is modeled as a manifold $\M$ equipped with a reference metric $\gbar = \gbar^0 + q$. A configuration is an embedding $\phi: \M \to \mathbb{R}^2$ from which an actual metric is defined on $\M$ as the pull-back of the euclidean metric on $\mathbb{R}^2$, denoted $g$.
We denoted by $\phi^*$ the energy minimizing configuration in the absence of external loads.
The equilibrium equations are derived from an energy variation with respect to the embedding $\phi$ describing the configuration.  
The elastic energy to be minimized is
\begin{eqnarray}
	F  =  \int_{\M} \W_\mathrm{el}(g,\gbar) \,\dif S_{\gbar} - \int_{\partial \M}  \mathbf{t} \cdot \mathbf{d} \,\dif l_{\gbar}\;.
	\label{eq:elasticenergy}
\end{eqnarray}
with $\mathbf{d} = \phi - \phi^*$, and
\begin{equation}
    \W_\mathrm{el}(g,\gbar) = \frac{1}{2} \A^{\alpha\beta\gamma\delta} \uel_{\alpha\beta}\uel_{\gamma\delta}\;.
\end{equation}
Upon defining the elastic stress 
\begin{eqnarray}
	\sigma_\mathrm{el}^{\alpha\beta} = \A^{\alpha\beta\gamma\delta} \uel_{\gamma\delta}
\end{eqnarray}
we find
\begin{eqnarray}
	\delta_\phi F  =  \int_{\M}\frac{1}{2}\sigma_\mathrm{el}^{\alpha\beta} \delta_\phi g_{\alpha\beta}  \,\dif S_{\gbar} -  \int_{\partial \M}  \mathbf{t} \cdot \delta \phi  \,\dif l_{\gbar}
	\;.
\end{eqnarray}
where we used $\delta \mathbf{d} = \delta (\phi - \phi^*) = \delta \phi$.

Writing the metric variation in terms of the configuration and using  $\delta_\phi g_{\alpha\beta} =  (\partial_\alpha \phi) (\partial_\beta \delta \phi) + (\partial_\alpha \delta \phi) (\partial_\beta \phi)$  we find 
\begin{eqnarray}
	\delta_\phi F  &=&  \int_{\M}\sigma_\mathrm{el}^{\alpha\beta} (\partial_{\alpha} \phi) (\partial_{\beta }\delta \phi)  \,\dif S_{\gbar} -  \int_{\partial \M}  \mathbf{t} \cdot \delta \phi  \,\dif l_{\gbar} \notag \\ 
	 &=& \oint_{\partial \M}\sigma_\mathrm{el}^{\alpha\beta} n_\beta (\partial_{\alpha} \phi) \delta \phi  \,\dif l_{\gbar}  \notag \\ &-& \int_{\M}\frac{1}{\sqrt{\gbar}}\partial_\beta \left(\sigma_\mathrm{el}^{\alpha\beta} (\partial_{\alpha} \phi)\sqrt{\gbar}\right) \delta \phi \,\dif S_{\gbar} \notag \\ &-&  \oint_{\partial \M}  \mathbf{t} \cdot \delta \phi  \,\dif l_{\gbar}
	\;.
\end{eqnarray}
In the second integral we note that integrand can be written as
\begin{eqnarray}
    \mathrm{div}_\beta \sigma_\mathrm{el}^{\alpha\beta} \partial_\alpha \phi &\equiv &\frac{1}{\sqrt{\gbar}}\partial_\beta \left(\sigma_\mathrm{el}^{\alpha\beta} (\partial_{\alpha} \phi)\sqrt{\gbar}\right)  \notag\\
    &=& \left({\nabla}_{\beta} \sigma_\mathrm{el}^{\alpha\beta} + \brk{\bar\Gamma^\nu_{\nu \beta}-{\Gamma}^\nu_{\nu \beta}} \sigma_\mathrm{el}^{\alpha \beta}  \right)\partial_\alpha \phi \notag
\end{eqnarray}
In the last integral we transform the vector $\mathbf{t}$ to the reference manifold by setting $\mathbf{t} = t^\mu \partial_{\mu} \phi$.  In this form the traction forces are defined on the reference manifold, which is equivalent to saying that the position on which forces applied are moving with the material, as in Lagrangian coordinates.
Therefore the variation takes the form
\begin{eqnarray}
	\delta_\phi F 
	 &=& \oint_{\partial \M} \left(\sigma_\mathrm{el}^{\alpha\beta} n_\beta  - t^\alpha \right) (\partial_{\alpha} \phi) \delta \phi  \,\dif l_{\gbar}  \notag \\ &-& \int_{\M}\mathrm{div}_\beta \sigma_\mathrm{el}^{\alpha\beta} \partial_\alpha \phi \,  \delta \phi \,\dif S_{\gbar} 
	\;.
\end{eqnarray}
We conclude that the equilibrium equation is
\begin{eqnarray}
	\bar{\nabla}_{\mu} \sigma_\mathrm{el}^{\mu\nu} + \brk{\Gamma^\nu_{\alpha \beta}-\bar{\Gamma}^\nu_{\alpha \beta}} \sigma_\mathrm{el}^{\alpha \beta} &=& 0,
\end{eqnarray}
along with the boundary conditions
\begin{equation}
	n_{\alpha} \sigma_\mathrm{el}^{\alpha \beta} = t^{\beta}\,.
\end{equation}

%%%%%%%%%%%%%%%%%%%%%%%%%%%%%%%%%%
%%%%%%%%%%%%%%%%%%%%%%%%%%%%%%%%%%
\section{Derivation of Equilibrium Equation for the induced quadrupoles}
\label{sec:IndQuadEq}
Here we derive the relation between the elastic stress and the induced quadrupoles in each screening regime. 

%%%%%%%%%%%%%%%%%%%%%%%%%%%%%%%%%%
\noindent \textbf{Quadrupole screening}\\
The variation of the work term \eqref{eq:QuadW}  with respect to $q$ yields
\begin{equation}
\begin{split}
    \int_{\M} \delta_q \W\,\dif S_{\gbar^0} = \int_{\M}
    \left( - {\Lambda}_{\mathrm{q}}^{\alpha\beta\gamma\delta} q_{\gamma\delta} \delta q_{\alpha\beta} \right) \,\dif S_{\gbar^0} 
	\; \sout{,} \Noemie{.}
\end{split}
\end{equation}
Substituting in \eqref{eq:VarFq} and requiring the variation to vanish we get
a linear relation between the induced quadrupole and the elastic stress
\begin{eqnarray}
	\sigmael^{\alpha\beta} + 2{\Lambda}_{\mathrm{q}}^{\alpha\beta\gamma\delta} q_{\gamma\delta} &=& 0\;.
	\label{eq:solquad}
\end{eqnarray}
Substituting the expressions for $q$ and $\Lambda_q$ in terms of $Q$ and $\Lambda^Q$ we obtain the first equation in \eqref{eq:ScreeningEq}.
%%%%%%%%%%%%%%%%%%%%%%%%%%%%%%%%%%%%%%%%%%%%%%%%%

\noindent \textbf{Dipole Screening}\\
The variation of the work term in the dipole screening regime reads
\begin{eqnarray}
	&&\int_{\M} \delta\W \, \dif S_{\gbar^0} = \notag\\
	&=&\int_{\partial \M}  \frac{1}{2} \lambda_P {\bar{\varepsilon}}^{\mu\alpha}{\bar{\varepsilon}}^{\nu\beta} \left( n_\mu P_\nu + n_\nu P_\mu\right)\delta q_{\alpha\beta}\,\dif l_{\gbar^0} \notag\\
	&-& \int_{\M}  \frac{1}{2} \lambda_P {\bar{\varepsilon}}^{\mu\alpha}{\bar{\varepsilon}}^{\nu\beta} \left( \bar\nabla_\mu P_\nu + \bar\nabla_\nu P_\mu\right) \delta q_{\alpha\beta}\, \dif S_{\gbar^0} 
	\; .
\end{eqnarray}
Substituting in \eqref{eq:VarFq} and requiring the total variation to vanish we obtain the second equation in \eqref{eq:ScreeningEq}.

\noindent \textbf{Monopole Screening}
To perform the variation in the monopole regime two integrations by parts are required. This is seen from the following:
\begin{eqnarray}
	\delta\W  
 % &=& \Lambda^{\mathrm{M}} (\bar{\nabla}_{\alpha\beta} Q^{\alpha\beta}) ( \bar{\nabla}_{\gamma\delta} \delta Q^{\gamma\delta}) \sqrt{\gbar_0} \notag \\ 
	% &=&\bar{\nabla}_{\gamma\delta}\left( \Lambda^{\mathrm{M}} (\bar{\nabla}_{\alpha\beta} Q^{\alpha\beta}) (  \delta Q^{\gamma\delta}) \sqrt{\gbar_0} \right)\notag \\
	% &-&\left( \Lambda^{\mathrm{M}} (\bar{\nabla}_{\gamma\delta} \bar{\nabla}_{\alpha\beta} Q^{\alpha\beta}) (  \delta Q^{\gamma\delta}) \sqrt{\gbar_0} \right)\notag \\
	% &-& 2 \left( \Lambda^{\mathrm{M}} (\bar{\nabla}_{\gamma}\bar{\nabla}_{\alpha\beta} Q^{\alpha\beta}) (  \delta \bar{\nabla}_{\delta}Q^{\gamma\delta}) \sqrt{\gbar_0} \right)\notag \\
	&=&\bar{\nabla}_{\gamma\delta}\left( \Lambda^{\mathrm{M}} (\bar{\nabla}_{\alpha\beta} Q^{\alpha\beta}) (  \delta Q^{\gamma\delta}) \sqrt{\gbar_0} \right)\notag \\
&-& 2 \bar{\nabla}_{\delta} \left( \Lambda^{\mathrm{M}} (\bar{\nabla}_{\gamma}\bar{\nabla}_{\alpha\beta} Q^{\alpha\beta}) (  \delta Q^{\gamma\delta}) \sqrt{\gbar_0} \right) \notag\\
&+& \left( \Lambda^{\mathrm{M}} (\bar{\nabla}_{\gamma\delta }\bar{\nabla}_{\alpha\beta} Q^{\alpha\beta}) (  \delta Q^{\gamma\delta}) \sqrt{\gbar_0} \right)
\; .
\end{eqnarray}
We substitute $M = \nablabar_{\alpha\beta}Q^{\alpha\beta}$, and note that the double integration by parts performed on the first term vanishes because $\partial\partial \M = 0$, that is the boundary of the boundary is closed. The variation therefore takes the form
\begin{eqnarray}
	\begin{split}
		\int_{\M} \delta\W \,  &\dif S_{\gbar^0} = \int_{\M}   \lambda_M {\bar{\varepsilon}}^{\mu\alpha}{\bar{\varepsilon}}^{\nu\beta} \left( \bar\nabla_{\mu \nu} M\right) \delta q_{\alpha\beta}\, \dif S_{\gbar^0}  \notag \\
		&- \int_{\partial \M}  \lambda_M {\bar{\varepsilon}}^{\mu\alpha}{\bar{\varepsilon}}^{\nu\beta} \left( n_\mu \bar\nabla_{\nu}M +n_\nu \bar\nabla_{\mu}M\right)\delta q_{\alpha\beta}\,\dif l_{\gbar^0}\; .
	\end{split}	
\end{eqnarray}
Substituting in \eqref{eq:VarFq} and requiring the total variation to vanish we obtain the third equation in \eqref{eq:ScreeningEq}.

\section{The normalized elastic tensor}
\label{app:RenormalizedA}
We firs relate $\Lambda^\mathrm{Q}$ with $\Lambda^\mathrm{q}$ as shown in \eqref{eq:QuadW}. 
Since $Q^{\alpha\beta} = \varepsilon^{\alpha\mu}\varepsilon^{\beta\nu} q_{\mu\nu}$ we find
\begin{eqnarray}
    \W_Q &=& \frac{1}{2} \Lambda^\mathrm{Q}_{\alpha\beta\gamma\delta} Q^{\alpha\beta}Q^{\gamma\delta} \\
    &=& \frac{1}{2} \Lambda^\mathrm{Q}_{\alpha\beta\gamma\delta} \varepsilon^{\alpha\mu}\varepsilon^{\beta\nu} q_{\mu\nu} \varepsilon^{\gamma\rho}\varepsilon^{\delta\sigma} q_{\rho\sigma} \\
    &\equiv& \frac{1}{2} \Lambda_\mathrm{q}^{\mu\nu\rho\sigma} q_{\mu\nu} q_{\rho\sigma}
\end{eqnarray}
with
\begin{eqnarray}
    \Lambda_\mathrm{q}^{\mu\nu\rho\sigma} =  \Lambda^\mathrm{Q}_{\alpha\beta\gamma\delta} \varepsilon^{\alpha\mu}\varepsilon^{\beta\nu} \varepsilon^{\gamma\rho}\varepsilon^{\delta\sigma}
\end{eqnarray}

Next we show how $\Lambda_\mathrm{q}$ normalizes the elastic tensor
in \eqref{eq:solquad} 
\begin{eqnarray}
		q_{\alpha\beta} = -\frac{1}{2}{\Lambda}^{\mathrm{q}}_{\alpha\beta\gamma\delta}   \sigmael^{\gamma\delta} \;.
\end{eqnarray}
Substituting in Eq.~\ref{eq:ScreenedConsRel} we find 
\begin{eqnarray}
	\sigmael^{\alpha\beta} &=& \A^{\alpha\beta\gamma\delta} u_{\gamma\delta} - \frac{1}{2} \A^{\alpha\beta\gamma\delta} q_{\gamma\delta}  \notag\\&=& \A^{\alpha\beta\gamma\delta} u_{\gamma\delta} - \frac{1}{2} \A^{\alpha\beta\gamma\delta} \left(-\frac{1}{2}{\Lambda}^{\mathrm{q}}_{\gamma\delta\mu\nu}   \sigmael^{\mu\nu}\right) 
\end{eqnarray}
Noting that 
\begin{eqnarray}
	\sigmael^{\alpha\beta} &=& \sigmael^{\mu\nu} \mathrm{Id}^{\alpha\beta}_{\quad \mu\nu} \notag \\ 
	\mathrm{Id}^{\alpha\beta}_{\quad \mu\nu}  &=& \frac{1}{2} \left( \delta^{\alpha}_{\,\,\mu}\delta^{\beta}_{\,\,\nu} + \delta^{\alpha}_{\,\,\nu}\delta^{\beta}_{\,\,\mu}\right) 
\end{eqnarray}
we get
\begin{eqnarray}
	\sigmael^{\mu\nu} \left(\mathrm{Id}^{\alpha\beta}_{\quad \mu\nu}  - \frac{1}{4} \A^{\alpha\beta\gamma\delta}{\Lambda}^{\mathrm{q}}_{\gamma\delta\mu\nu} \right) &=& \A^{\alpha\beta\gamma\delta} u_{\gamma\delta}\;.
\end{eqnarray}
Upon denoting 
\begin{eqnarray}
	\Gamma^{\alpha\beta}_{\quad \mu\nu} = \mathrm{Id}^{\alpha\beta}_{\quad \mu\nu}  - \frac{1}{4} \A^{\alpha\beta\gamma\delta}{\Lambda}^{\mathrm{q}}_{\gamma\delta\mu\nu}
\end{eqnarray}
we get 
\begin{eqnarray}
	\sigmael^{\mu\nu} &=& {\Gamma^{-1}}_{\alpha\beta}^{\quad\mu\nu} \A^{\alpha\beta\gamma\delta} u_{\gamma\delta}\;.
\end{eqnarray}
that is 
\begin{eqnarray}
	 \tilde\A^{\alpha\beta\gamma\delta} = {\Gamma^{-1}}_{\mu\nu}^{\quad\alpha\beta} \A^{\mu\nu\gamma\delta}
\end{eqnarray}

Note that in the absence of quadrupole screening, where all the coefficients in Eq.~\ref{eq:cost} vanishes, $\Gamma$ reduces to the identity, and the elastic tensor remains intact.

%%%%%%%%%%%%%%%%%%%%%%%%%%%%%%%%%%%%%%%%
\section{Derivation of Induced Effective Monopoles}
\label{app:InducedMonopoles}
To derive the induced monopole charge distribution $M_\mathrm{ind} = \nablabar_{\alpha\beta}Q^{\alpha\beta}$ we use the relation between stress and induced quadrupoles given in \eqref{eq:ScreeningEq}. 
In the quadrupole, in the first equation in \eqref{eq:ScreeningEq}, we take the second divergence to express $\nablabar_{\alpha\beta}Q^{\alpha\beta}$. The divergence of the elastic stress, and therefore its second divergence as well, vanishes in equilibrium, hence in this regime $\nablabar_{\alpha\beta}Q^{\alpha\beta} = 0$.

In the dipole screening regime we take the trace of the second equation in \eqref{eq:ScreeningEq} and find 
\begin{equation}
    \mathrm{Tr} \, \sigmael = -Y \ellp^2 \gbar^0_{\alpha\beta} {\bar{\varepsilon}}^{\mu\alpha}{\bar{\varepsilon}}^{\nu\beta} \left( \bar\nabla_\mu P_\nu + \bar\nabla_\nu P_\mu\right)
\end{equation}
Upon substituting $P$ in terms of $Q$ and $\mathrm{Tr} \, \sigmael = \Deltabar \chi$ we obtain the second equation in \eqref{eq:IndMon}
\begin{eqnarray}
    \nablabar_{\alpha\beta}Q^{\alpha\beta} = -\frac{1}{2 Y \ellp^2 }  \bar{\Delta} \chi
\end{eqnarray}

% For the monopole case we rewrote the third equation in \eqref{eq:ScreeningEq} in the form
% \begin{eqnarray}
% 	\bar{\varepsilon}^{\alpha\mu}\bar{\varepsilon}^{\beta\nu} \bar{\nabla}_{\mu\nu} \left( \chi + \Lambda^M  \bar{\nabla}_{\alpha\beta} Q^{\alpha\beta} \right) = 0 \;. 
% \end{eqnarray}
% and concluded that 
% \begin{eqnarray}
%     \chi + \Lambda^M  \bar{\nabla}_{\alpha\beta} Q^{\alpha\beta}  = \chi_g \;, 
% \end{eqnarray}
% where $\chi_g$ is any function satisfying $\bar{\nabla}_{\mu\nu} \chi_g = 0$, reflecting the gauge freedom of the stress function.

Lastly, for monopole screening regime, substituting \eqref{eq:HexSolRep} in \eqref{eq:QuadMon} we find
\begin{eqnarray}
	\bar{\varepsilon}^{\alpha\mu}\bar{\varepsilon}^{\beta\nu} \bar{\nabla}_{\mu\nu} \left( \chi + Y \ell_M^4  \bar{\nabla}_{\alpha\beta} Q^{\alpha\beta} \right) = 0 \;. 
\end{eqnarray}
We conclude that 
\begin{eqnarray}
    \chi + Y \ell_M^4  \bar{\nabla}_{\alpha\beta} Q^{\alpha\beta}  = \chi_g \;, 
\end{eqnarray}
where $\chi_g$ is any function satisfying $\bar{\nabla}_{\mu\nu} \chi_g = 0$, reflecting the gauge freedom of the stress function. Upon setting a gauge such that $\chi_g = 0$ we find
\begin{eqnarray}
      \bar{\nabla}_{\alpha\beta} Q^{\alpha\beta}  = -\frac{1}{Y \ell_M^4}\chi \;. 
      \label{eq:MonGaugeChoice}
\end{eqnarray}
%%%%%%%%%%%%%%%%%%%%%%%%%%%%%%%%%%%%%%%%
\section{Interactions}
\label{app:Int}
In this section we derive the interaction-form of the mechanical energy stored in the screened solid. The case of quadrupole screening require no analysis since the only effect of the induced quadrupoles is normalizing the elastic tensor and the interaction rmeain intact apart from normalized elastic moduli. 

In the case of dipole screening, the total energy is 
\begin{eqnarray}
    E &=&  \int_{\M} \left( \frac{1}{2} \A^{\alpha\beta\gamma\delta} \uel_{\alpha\beta}\uel_{\gamma\delta}  -  \frac{1}{2} \Lambda^{\mathrm{P}}_{\alpha\beta} P^\alpha P^\beta\right) \,\dif S_{\gbar^0} \notag\\
   &=&  \int_{\M} \left( \frac{1}{2} \sigma_\mathrm{el}^{\alpha\beta}\uel_{\alpha\beta} -  \frac{1}{2} \lambda_P \nablabar_\mu Q^{\mu\alpha} P_\alpha\right) \,\dif S_{\gbar^0} \notag\\
   &=&  \int_{\M} \left( \frac{1}{2} \sigma_\mathrm{el}^{\alpha\beta}u_{\alpha\beta} -  \frac{1}{4} \sigma_\mathrm{el}^{\alpha\beta}q_{\alpha\beta} +  \frac{1}{2} \lambda_P  Q^{\mu\alpha} \nablabar_{\mu} P_{\alpha}\right) \,\dif S_{\gbar^0} \notag \\
    &-& \int_{\partial \M} \lambda_P Q^{\mu\alpha} P_\alpha n_\mu \dif l_{\gbar^0}
\end{eqnarray}
 Using the symmetry of $Q$ and substituting it in terms of $q$ we find that the boundary term vanishes from the boundary condition in \eqref{eq:DipoleBC}, and the second and third terms in the integral cancel from the equilibrium equation  \eqref{eq:DipoleEquilibrium}.
 We therefore conclude
 \begin{eqnarray}
    E &=&  \int_{\M}  \frac{1}{2} \sigma_\mathrm{el}^{\alpha\beta}u_{\alpha\beta} \,\dif S_{\gbar^0} 
\end{eqnarray}
Upon expressing $\sigmael$ in terms of the stress function and integrating by parts twice we find that at the linear approximation
 \begin{eqnarray}
    E &=&  \int_{\M}  \frac{1}{2} \sigma_\mathrm{el}^{\alpha\beta}u_{\alpha\beta} \,\dif S_{\gbar^0} = \int_{\M}  \chi \Kbar \,\dif S_{\gbar^0}\;.
\end{eqnarray}

\section{Complete solution for Green's function}
\label{app:2}
The Green's function within the screened elasticity setup is the solution for \eqref{eq:ScreenedSF} with a delta-function singularity, as solved in \eqref{eq:GreensDipole} and \eqref{eq:GreensMonopole}.
The solution is first derived for a finite domain with traction free boundary conditions. In the case of dipole screening the constants of integration are
\begin{eqnarray}
     c_1 &=& \frac{q}{2 \pi \rin \rout} \frac{\rin Y_1\left(\rin\right)-\rout Y_1\left(\rout\right)}{Y_1\left(\rin\right) J_1\left(\rout\right)-J_1\left(\rin\right) Y_1\left(\rout\right)}\notag\\
     c_2 &=& \frac{q}{2 \pi \rin \rout} \frac{\rin J_1\left(\rin\right)-\rout J_1\left(\rout\right)}{J_1\left(\rin\right) Y_1\left(\rout\right)-Y_1\left(\rin\right) J_1\left(\rout\right)}\notag
\end{eqnarray}
and here $\rin$ and $\rout$ are measured in units of $r_s = \sqrt{2\Lambda_P}$. In the limit $\rout \to \infty$ we both constants vanish, leading to the Green's function $G^\text{DS}$ given in \eqref{eq:DipScGreen}.

%%%%%%%%%%%%%%%%%%%%%%%%%%%%%%%%%%%%%
%%%%%%%%%%%%%%%%%%%%%%%%%%%%%%%%%%%%%

\bibliography{refs}

\end{document}